\documentstyle[12pt]{article}
\begin{document}

\renewcommand{\thesection}{\arabic{section}}
\renewcommand{\theequation}{\thesection .\arabic{equation}}
\newcommand{\sect}[1]{\section{#1}\setcounter{equation}{0}}
\newcommand{\subsect}[1]{\subsection{#1}}
\newcommand{\alphsec}{\setcounter{section}{0}
                      \renewcommand{\thesection}{\Alph{section}}}
\newcommand{\appe}[1]{\setcounter{equation}{0}
                      \setcounter{subsection}{0}
                      \addtocounter{section}{1}
                      \section*{Appendix \ #1}}
\newcommand{\beq}{\begin{equation}}
\newcommand{\eeq}{\end{equation}}
\newcommand{\beqs}{\begin{eqnarray}}
\newcommand{\eeqs}{\end{eqnarray}}

\begin{titlepage}

\null

\begin{flushright}
 NBI-HE-93-01 \\
 KOBE-TH-93-01 \\
 January 1993
\end{flushright}

\vspace{1cm}
\begin{center}
 {\Large\bf
  Topological Aspects of \par}
\vspace{6mm}
 {\Large\bf
  an Antisymmetric Background Field on Orbifolds
  \par}
\vspace{2.5cm}
\baselineskip=7mm
 {\large
  Makoto Sakamoto
  \footnote{On leave from Department of Physics,
            Kobe University, Rokkodai, Nada, Kobe 657, Japan.}
  \par}
 {\sl
  The Niels Bohr Institute, University of Copenhagen\\
  Blegdamsvej 17, DK-2100 Copenhagen X, Denmark
  \par}

\vspace{3cm}
 {\large\bf Abstract}
\end{center}
\par

We study string theory on orbifolds in the presence of an antisymmetric
constant background field in detail and discuss some of new aspects of
the theory.
It is pointed out that the term with the antisymmetric background
field can be regarded as a topological term like a Chern-Simons term
or a Wess-Zumino term.
Detailed analysis in the operator formalism
shows that orbifold models with topologically nontrivial twists
exhibit various anomalous behavior:
Zero mode variables obey nontrivial quantization conditions.
Coordinate transformations which define orbifolds are modified
at quantum level.
Wave functions of twisted strings in general acquire nontrivial
phases when they move around non-contractible loops on orbifolds.
Zero mode eigenvalues are shifted from naively expected values,
in favor of modular invariance.

\end{titlepage}

\baselineskip=7mm

%%%%%%%%%%%%%%%%%%%%%%%%%%%%%%%%%%%%%%%%%%%%%%%%%%%%%%%%%%%%%%%%%%
%%%%%%%%%%%%%%%%%%%%%%%%%%%%%%%%%%%%%%%%%%%%%%%%%%%%%%%%%%%%%%%%%%

\sect{Introduction}

Our understanding of geometry of string compactifications
is far from complete even for simple compactifications.
For instance, in toroidal compactification there appears a degree
of freedom of a winding number.
Since a winding number of a closed string can change in string
interactions, we need to introduce a canonical \lq\lq coordinate"
$ Q^I $ conjugate to the winding number
in order to construct the quantum theory.
Physical roles of the \lq\lq coordinate" $Q^I$ are clear but
geometrical meanings of $Q^I$ are not obvious.
This fact may cause a problem of how to impose a quantization
condition on $Q^I$.
We will face this problem in our discussions and see that the
consistency with string interactions requires a nontrivial
quantization condition on $Q^I$.

Orbifold compactification
\cite{orbifold} is also as simple as toroidal one and
is probably the most efficient method to construct
a realistic four dimensional string model.
On orbifolds there also appears the \lq\lq coordinate" $Q^I$.
The lack of geometrical meanings of $Q^I$ may cause another problem
of how to determine the action of twist operators on $Q^I$.
We will see that our naive intuition fails in some cases.

Our aim of this paper is to discuss some of new
aspects of string
theory on orbifolds in the presence of an antisymmetric background
field.
Orbifold models have been discussed by many authors and various
phenomenologically \lq\lq realistic" models have been proposed
\cite{orbmodels}.
Nevertheless, one interesting topological nature of the antisymmetric
background field has not been discussed yet.

A $D$-dimensional torus $T^D$ is defined by identifying
a point \{$X^I$\} to
\{$X^{I}+\pi w^I$\} for all $w^I \in \Lambda$, where $\Lambda$ is
a $D$-dimensional lattice.
An orbifold is obtained by dividing a torus by the action of a
discrete symmetry group $G$ of the torus.
Any element $g$ of $G$ can be represented (for symmetric orbifolds) by
\cite{orbifold,Sakamoto-TH}
%%%%%%%%%%%%%%%%%%%%% 1.1
\begin{equation}
g\ =\ (\ U\ ,\ v\ )\ \ ,
\end{equation}
%%%%%%%%%%%%%%%%%%%%%
where $U$ denotes a rotation and $v$ a translation.
In sect. 2, we clarify a topological nature of the term with the
antisymmetric background field $B^{IJ}$:
Consider a transformation,
%%%%%%%%%%%%%%%%%%%%% 1.2
\begin{equation}
g\ :\quad X^I\ \ \rightarrow \ \ U^{IJ}X^J\ .
\label{XtoUX}
\end{equation}
%%%%%%%%%%%%%%%%%%%%%%
It is shown in the next section that a Euclidean action $S_{E}[X]$
in the path integral formalism is not invariant under the
transformation (\ref{XtoUX}) unless the rotation matrix $U^{IJ}$
commutes with $B^{IJ}$ but in general
%%%%%%%%%%%%%%%%%%%%% 1.3
\begin{equation}
S_{E}[X]\ -\ S_{E}[UX]\ =\ i2\pi n\quad\mbox{ with\ $n \in
{\bf Z}$}\ .
\label{CS}
\end{equation}
%%%%%%%%%%%%%%%%%%%%%%%
This means that the transformation (\ref{XtoUX}) is not a symmetry
of the Euclidean action $S_{E}[X]$ unless $[U,B]=0$ but is
a symmetry of the theory at the quantum level.
Thus the term with the antisymmetric background field can be regarded
as a topological term like a Chern-Simons term or
a Wess-Zumino term. Most attention of previous works has been paid
to the case
%%%%%%%%%%%%%%%%%%%%%%%%
\begin{equation}
[\ U\ ,\ B\ ]\ =\ 0\ ,
\end{equation}
%%%%%%%%%%%%%%%%%%%%%%%%%
or simply $B^{IJ}=0$, i.e. to
the case $n=0$ in eq. (\ref{CS}).
The generalization to the case
%%%%%%%%%%%%%%%%%%%%%%%
\begin{equation}
[\ U\ ,\ B\ ]\ \ne\ 0\ ,
\label{nontrivial}
\end{equation}
%%%%%%%%%%%%%%%%%%%%%%
is not, however, trivial, as expected from the result (\ref{CS}).
A naive construction of orbifold models with eq.(\ref{nontrivial})
would give modular non-invariant partition functions and also destroy
the duality of amplitudes
\cite{Sakamoto-TH,Inoue-NT,Erler-JLM}.
Here, we will call the twist $g$ in eq. (\ref{XtoUX})
topologically trivial (nontrivial) if $[U,B]=0$ ($[U,B]\ne 0$).
Our concern of this paper is orbifold models with topologically
nontrivial twists $g$.
In ref.
\cite{Sakamoto92},
some of the properties of such orbifold models
have been reported. In this paper, we will give
the full details.

In sect. 3, we summarize the basics of strings on tori in the
operator formalism and then derive consistency conditions
for constructing string theory on orbifolds.

Due to the property (\ref{CS}), orbifold models with eq.
(\ref{nontrivial}) exhibit various anomalous behavior.
In sect. 4, we discuss the cocycle property of vertex operators
in detail and show that zero modes of strings should obey
nontrivial quantization conditions, which are the origin of the
anomalous behavior of the theory in an operator formalism
point of view.
We also show that coordinate transformations become anomalous
for topologically nontrivial twists and that a vertex operator
$V(k_L,k_R;z)$ transforms as
%%%%%%%
\beq
g:\quad V(k_L,k_R;z)\ \rightarrow\ \rho\ V(U^Tk_L,U^Tk_R;z)\ .
\eeq
%%%%%%%%%%%%%%%%
A nontrivial phase $\rho$ may appear for a topologically
nontrivial twist $g$ and can be regarded as a kind of quantum
effects.
The phase $\rho$ plays an important role in extracting
physical states.

In sect. 5, we construct the complete set and the inner product
of zero mode eigenstates.
The construction is not straightforward due to nontrivial
quantization conditions of zero modes.
We find that zero mode eigenvalues differ from naively
expected ones in (topologically nontrivial) twisted sectors.
We also find Aharonov-Bohm like effects
\cite{ABeffect} in our system.
The antisymmetric background field plays a similar role of an
external gauge field.
A wave function of a twisted string is not, in general, periodic
with respect to torus shifts but
%%%%%%% 1.7
\beq
\Psi(x^I+\pi w^I)\ =\ e^{-i\pi w^Is^I_g}\ \Psi(x^I)\ ,
\eeq
%%%%%%%%%%%%%%%
for $w^I \in \Lambda$ such that $w^I=U^{IJ}w^J.$
The constant vector $s^I_g$ depends on the commutator $[U,B].$

In sect. 6, we compute one loop partition functions and prove
modular invariance of them.
This justifies our formulation.

In sect. 7, we present an example of an orbifold model with
a topologically nontrivial twist, which may give a good
illustration of our formulation.
Sect. 8 is devoted to discussion.

In appendix A, we give a precise definition of the vertex operator
and derive a consistency condition which is required from the
duality of amplitudes and which is a key equation of our discussions.
In appendix B, we present various useful formulas of the Jacobi
theta functions.
In appendix C, we derive equations (eqs. (\ref{pathint}) and
(\ref{pathint'}) below)
which we will use to prove modular invariance in sect. 6.

\vspace{1cm}

%%%%%%%%%%%%%%%%%%%%%%%%%%%%%%%%%%%%%%%%%%%%%%%%%%%%%%%%%%%%%%%%%%%%%%
%%%%%%%%%%%%%%%%%%%%%%%%%%%%%%%%%%%%%%%%%%%%%%%%%%%%%%%%%%%%%%%%%%%%%%

\sect{A Topological Nature of the Antisymmetric
         Background Field}

In this section, we shall clarify a topological nature of the
antisymmetric background field from a path integral
point of view.
Since fermionic degrees of freedom play no important role in
our discussions, we will restrict our considerations to
bosonic strings.

In the Euclidean path integral formalism, the one-loop vacuum amplitude
of the closed bosonic string theory on a torus is given by the functional
integral
\cite{Polyakov-P},
%%%%%%%%%%%%%%%%%%%%%%% 2.1
\begin{equation}
\int {[dg_{\alpha \beta}][dX^I]\over{\cal V}}\  \exp \{ -S_E[X,g] \} ,
\label{epath}
\end{equation}
%%%%%%%%%%%%%%%%%%%%%%
where $g_{\alpha \beta}$ is a Euclidean metric of the two dimensional
world sheet with the topology of a torus and ${\cal V}$ is a volume
of local gauge symmetry groups. The Euclidean action is given by
%%%%%%%%%%%%%%%%%%%%%%%% 2.2
\begin{equation}
S_E[X,g]=\int_0^\pi d^2\sigma {1\over2\pi}
\left\{\sqrt g g^{\alpha \beta}
\partial _{\alpha} X^I \partial_{\beta}X^I
-iB^{IJ}\varepsilon^{\alpha
\beta}\partial_{\alpha}X^I \partial_{\beta}X^J\right\}\ ,
\label{eaction}
\end{equation}
%%%%%%%%%%%%%%%%%%%%%%%%
where $B^{IJ}\ (I,J=1,\cdots,D)$ is the antisymmetric constant
background field and $X^I\ (I=1,\cdots,D)$ is a string coordinate.
It should be emphasized that the imaginary number $i$ appears in the
second term of the Euclidean action (\ref{eaction})
due to the antisymmetric
property  of $\varepsilon^{\alpha \beta}$. Since strings propagate on
a torus which is defined by identifying a point \{$X^I$\} with
\{$X^I+\pi w^I$\} for
all $w^I \in \Lambda$, where $\Lambda$ is a $D$-dimensional lattice,
the string coordinate $X^I(\sigma^1,\sigma^2)$ obeys the following
boundary conditions:
%%%%%%%%%%%%%%%%%%%%%%% 2.3
\begin{eqnarray}
   X^I(\sigma^1+\pi,\sigma^2) &=& X^I(\sigma^1,\sigma^2) + \pi w^I\ ,
\nonumber\\
   X^I(\sigma^1,\sigma^2+\pi) &=& X^I(\sigma^1,\sigma^2) + \pi{w'}^I\ ,
\label{pathbc}
\end{eqnarray}
%%%%%%%%%%%%%%%%%%%%%%% 2.4
for some $w^I,{w'}^I \in \Lambda$.

Let us consider a transformation,
%%%%%%%%%%%%%%%%%%%%%
\begin{equation}
g :\quad X^I\  \longrightarrow \ U^{IJ}X^J \quad
\mbox{with}\ \
U^TU = \mbox{\bf 1}\ ,
\label{XtoUX2}
\end{equation}
%%%%%%%%%%%%%%%%%%%%%%%
where $U^{IJ}$ is an orthogonal matrix. The first term of the action
(\ref{eaction}) is trivially invariant under the transformation
(\ref{XtoUX2}) but the
second term is not invariant if $U^{IJ}$ does not commute with $B^{IJ}$.
Since $B^{IJ}$ is a constant antisymmetric field, the second term of the
action (\ref{eaction}) can be written as a total divergence, i.e.
%%%%%%%%%%%%%%%%%%%% 2.5
\begin{equation}
B^{IJ} \varepsilon^{\alpha \beta}\partial_\alpha X^I
\partial_\beta X^J = \partial_\alpha(B^{IJ}\varepsilon^{\alpha \beta}
X^I \partial_\beta X^J)\ .
\end{equation}
%%%%%%%%%%%%%%%%%%%%%%
It turns out that the difference between $S_E[X,g]$ and $S_E[UX,g]$ is
given by
%%%%%%%%%%%%%%%%%%%%%% 2.6
\begin{equation}
S_E[X,g] - S_E[UX,g] = i\pi {w'}^I (B-U^TBU)^{IJ}w^J\ .
\end{equation}
%%%%%%%%%%%%%%%%%%%%%%
{}From this relation, we come to an important conclusion that the
transformation (\ref{XtoUX2}) is a {\em quantum\/} symmetry of the
theory in a path integral point of view if
%%%%%%%%%%%%%%%%%%%%%% 2.7
\begin{equation}
(B-U^TBU)^{IJ}w^J \in 2\Lambda^* \quad \mbox{for all}\ \
w^I \in \Lambda\ ,
\label{consist}
\end{equation}
%%%%%%%%%%%%%%%%%%%%%%%
where $\Lambda^*$ is the dual lattice of $\Lambda$, although the
Euclidean action (\ref{eaction}) itself is not invariant under
the transformation
(\ref{XtoUX2}) if $[U,B]\ne 0$.
Then, the second term of the action (\ref{eaction})
can be regarded as a topological term like a Chern-Simons term or a
Wess-Zumino term, as announced before.
We will show in the next section that the condition (\ref{consist})
is nothing but a consistency condition for constructing string
theory on orbifolds.

\vspace{1cm}

%%%%%%%%%%%%%%%%%%%%%%%%%%%%%%%%%%%%%%%%%%%%%%%%%%%%%%%%%%%%%%%%%%%%%%%
%%%%%%%%%%%%%%%%%%%%%%%%%%%%%%%%%%%%%%%%%%%%%%%%%%%%%%%%%%%%%%%%%%%%%%

\sect{Consistency Conditions for Constructing String Theory
       on Orbifolds}

In order to fix the notation, we will first summarize the basics of
strings on tori in the operator formalism
and then derive consistency conditions for constructing
string theory on orbifolds.

In this paper, we shall study string theory on orbifolds in the
presence of the antisymmetric constant background field in detail.
The antisymmetric background field has been introduced by
Narain, Sarmadi and Witten
\cite{Narain2} to explain Narain torus
compactification
\cite{Narain1} in the conventional approach.
Narain torus compactification will be described in terms of
left- and right-moving coordinates and momenta, while in the
conventional approach we will use a winding number and a center of
mass coordinate and momentum.
The combined left-right momentum
of Narain torus compactification forms a Lorentzian even self-dual
lattice and this fact can be interpreted physically as tuning
on the antisymmetric background field on toroidally compactified
string theories.
The use of the left- and right-moving variables may be convenient
to discuss the mathematical structure of string theory on tori
but the geometrical interpretation is less clear.
Due to the lack of geometrical meanings of the left- and right-
moving variables,
we might have trouble quantizing string theory on orbifolds
based on Narain torus compactification, in particular, quantizing
twisted strings.
Hence, we will follow the conventional approach in this paper
although we may also use the left- and right-moving variables
for convenience.

In the construction of an orbifold model,
we will start with a D-dimensional toroidally compactified
string theory.
The action from which we shall start is
%%%%%%%%%%%%%%%%% 3.1
\begin{equation}
S[X]=
 \int d\tau\int_0^\pi d\sigma {1\over2\pi}
 \left\{\eta^{\alpha \beta}
 \partial _{\alpha} X^I(\tau,\sigma) \partial_{\beta}X^I(\tau,\sigma)
 +B^{IJ}\varepsilon^{\alpha
 \beta}\partial_{\alpha}X^I(\tau,\sigma)
 \partial_{\beta}X^J(\tau,\sigma)\right\}\ ,
\label{maction}
\end{equation}
%%%%%%%%%%%%%%%%%%%%%%%%
where $\eta^{\alpha\beta}$ is a two-dimensional Minkowski metric
and $\varepsilon^{01}=-\varepsilon^{10}=1.$
The $X^I(\tau,\sigma)$ $(I=1,\cdots,D)$ is a string coordinate on the
$D$-dimensional torus $T^D$ which is defined by identifying
a point \{$X^I$\} with
\{$X^I+\pi w^I$\} for all $w^I \in \Lambda$, where $\Lambda$ is a
$D$-dimensional lattice.
Since the second term in eq.(\ref{maction}) is a total divergence, it
does not affect the equation of motion.
The canonical momentum conjugate to $X^I(\tau,\sigma)$,
however, becomes
%%%%%%%%%%%%%%%%%%%%%% 3.2
\begin{equation}
P^I(\tau,\sigma)\ =\ \frac1{\pi}\Bigl(\partial_\tau X^I(\tau,\sigma)
\ +\ B^{IJ}\partial_\sigma X^J(\tau,\sigma)\Bigr)\ .
\end{equation}
%%%%%%%%%%%%%%%%%%%%%%%
Therefore, the mode expansion of $X^I(\tau,\sigma)$ is given by
%%%%%%%%%%%%%%%%%%%%%% 3.3
\begin{equation}
X^I(\tau,\sigma)\ =\ x^I\ +\ (p^I - B^{IJ}w^J)\tau\ +\
w^I\sigma\ +\ (\mbox{oscillators})\ ,
\end{equation}
%%%%%%%%%%%%%%%%%%%%%%%%
where $p^I$ is the canonical momentum conjugate to the center of mass
coordinate $x^I$ and $w^I$ is the winding number.
The string coordinate $X^I(\tau,\sigma)$ obeys the boundary
condition
%%%%%%%%%%%%%%%%%%%%%  3.4
\begin{equation}
X^I(\tau,\sigma+\pi)\ =\ X^I(\tau,\sigma)\ +\ \pi w^I\ ,
\end{equation}
%%%%%%%%%%%%%%%%%%%%%
where $w^I \in \Lambda$.

In order to construct the quantum theory of strings on tori, we need to
introduce the canonical \lq\lq coordinate" $Q^I$ conjugate to $w^I$
because the winding number $w^I$ appears in the spectrum and
string interactions can change the winding number.
We assume that the following canonical commutation relations:
%%%%%%%%%%%%%%%%% 3.5
\begin{eqnarray}
  [\ x^I\ ,\ p^J\ ]\ &=&\ i\delta^{IJ}\ , \nonumber\\
  {[\ Q^I\ ,\ w^J\ ]}\ &=&\ i\delta^{IJ}\ .
\end{eqnarray}
%%%%%%%%%%%%%%%%%%%
Since $x^I$ is the center of mass coordinate of a string moving
on the torus $T^D$, the wave function $\Psi(x^I)$ must be periodic,
i.e. $\Psi(x^I+\pi w^I)=\Psi(x^I)$ for all $w^I \in \Lambda$.
Hence, the allowed momentum is
%%%%%%%%%%%%%%%%3.6
\begin{equation}
p^I\ \in \ 2\Lambda^{*}\ ,
\end{equation}
%%%%%%%%%%%%%%%%%
where $\Lambda^{*}$ is the dual lattice of $\Lambda$.

For later convenience, we introduce the left- and right-moving
coordinates,
%%%%%%%%
\beq
X^I(\tau,\sigma) = \frac{1}{2}\left( X^I_L(\tau+\sigma) +
                   X^I_R(\tau-\sigma) \right)\ ,
\eeq
%%%%%%%%%%%%%%%
where
%%%%%%%
\beqs
X^I_L(\tau+\sigma) = x^I_L + 2p^I_L(\tau+\sigma)
  + i \sum_{n\ne 0}\frac{1}{n}\alpha^I_{Ln} e^{-2in(\tau+\sigma)}\ ,
   \nonumber\\
X^I_R(\tau-\sigma) = x^I_R + 2p^I_R(\tau-\sigma)
  + i \sum_{n\ne 0}\frac{1}{n}\alpha^I_{Rn} e^{-2in(\tau-\sigma)}\ .
\eeqs
%%%%%%%%%%%%%
The relations between $x^I_L,x^I_R,p^I_L,p^I_R$ and
$x^I,Q^I,p^I,w^I$ are given by
%%%%%%%%%%%%%%%%%% 3.7
\begin{eqnarray}
x^I_L&=&(1-B)^{IJ}x^J+Q^I\ ,\nonumber\\
x^I_R&=&(1+B)^{IJ}x^J-Q^I\ ,\nonumber\\
p^I_L&=&\frac1{2}(p^I-B^{IJ}w^J)+\frac1{2}w^I\ ,\nonumber\\
p^I_R&=&\frac1{2}(p^I-B^{IJ}w^J)-\frac1{2}w^I\ ,
\label{relation1}
\end{eqnarray}
%%%%%%%%%%%%%%%%%% 3.8
or equivalently,
%%%%%%%%%%%%%%%%%
\begin{eqnarray}
x^I&=&\frac1{2}(x^I_L+x^I_R)\ ,\nonumber\\
Q^I&=&\frac1{2}(1+B)^{IJ}x^J_L-\frac1{2}(1-B)^{IJ}x^J_R\ ,
    \nonumber\\
p^I&=&(1+B)^{IJ}p^J_L+(1-B)^{IJ}p^J_R\ ,\nonumber\\
w^I&=&p^I_L-p^I_R\ .
\label{relation2}
\end{eqnarray}
%%%%%%%%%%%%%%%%%%%
It follows from the definition (\ref{relation1}) or (\ref{relation2})
that the left- and right-moving momentum $(p^I_L,p^I_R)$
forms a $(D+D)$-dimensional Lorentzian even self-dual lattice
$\Gamma^{D,D}$
\cite{Narain1,Narain2},
%%%%%%%%%%%%%%%%%
\begin{equation}
(\ p^I_L\ ,\ p^I_R\ ) \in \Gamma^{D,D}\ .
\end{equation}
%%%%%%%%%%%%%%%
This observation is important to one loop modular invariance.

It is worthwhile noting that the following replacement of the
antisymmetric background field does not change the spectrum
\cite{axionicshift}:
%%%%%%%%%%%%%%
\beq
B^{IJ}\ \ \rightarrow\ \ B^{IJ}\ +\ \Delta B^{IJ}\ ,
\eeq
%%%%%%%%%%%%%%%
if $\Delta B^{IJ}w^J \in 2\Lambda^{*}$ for all $w^I \in \Lambda.$
This is because the momentum $p^I$ always appears in the
combination $p^I - B^{IJ} w^J$ in eq.(\ref{relation1}).
This discrete symmetry is the origin of the symmetry discussed in
sect. 2.

An orbifold is obtained by dividing a torus by the action of a suitable
discrete group $G$. In other words, an orbifold is given by the
identification of a point $\{X^I\}$  with $\{g\cdot X^I + \pi w^I\}$
for all $g \in G$ and $w^I \in \Lambda$.
In general any element $g$ of $G$ can be represented (for symmetric
orbifolds) by
%%%%%%%%%%%%%%%%%%
\beq
g\ =\ (\ U\ ,\ v\ )\ ,
\eeq
%%%%%%%%%%%%%%%%%%
where $U$ denotes a rotation and $v$ is a translation.
Since the shift vector $v$ is irrelevant in our discussions, we
will set $v=0$ throughout this paper for simplicity.
Since a point $\{X^I\}$ is identified with $\{U^{IJ}X^J\}$ up to
a torus shift on the orbifold, the action of $g$ on the string
coordinate must be well-defined,
%%%%%%%%%%%%%%%
\beq
g\ :\quad X^I(\tau,\sigma)\ \rightarrow\ U^{IJ}X^J(\tau,\sigma)\ .
\label{map}
\eeq
%%%%%%%%%%%%%%%%%
In terms of the zero modes, the map (\ref{map}) can be written as
%%%%%%%%%%%%%%%%
\beqs
g\ :\quad x^I\ &\rightarrow&\ U^{IJ}x^J\ ,\nonumber\\
          p^I\ &\rightarrow&\ U^{IJ}p^J\ -\ [U,B]^{IJ}w^J\ ,
             \nonumber\\
          w^I\ &\rightarrow&\ U^{IJ}w^J\ .
\eeqs
%%%%%%%%%%%%%%%%%%
Since the winding number $w^I$ lies on the lattice $\Lambda$,
$g$ must be an automorphism of the lattice $\Lambda$, i.e.
%%%%%%%%%%%%%%%%
\beq
g\ :\quad \Lambda\ \rightarrow \ \Lambda \qquad (\mbox{or}
\ \Lambda^{*} \rightarrow \Lambda^{*})\ .
\label{consist1}
\eeq
%%%%%%%%%%%%%%%%
Since the center of mass momentum $p^I$ lies on the lattice
$2\Lambda^{*}$, we further need to require the condition,
%%%%%%%%%%%%%
\begin{eqnarray*}
[\ U\ ,\ B\ ]^{IJ}w^J\ \in\ 2\Lambda^{*}\ ,
\end{eqnarray*}
%%%%%%%%%%%%%%%
or equivalently,
%%%%%%%%%%%%%%%
\beq
(B-U^{T}BU)^{IJ}w^J\ \in\ 2\Lambda^{*}\qquad
\mbox{for all}\ w^I\ \in\ \Lambda\ .
\label{consist2}
\eeq
%%%%%%%%%%%%%%%%
This is just the condition (\ref{consist}), as announced before.

In terms of the left- and right-moving momentum, the action of $g$
is given by
%%%%%%%%%%%%%% 3.17
\beq
g\ :\quad (p^I_L,p^I_R)\ \rightarrow\ (U^{IJ}p^J_L,U^{IJ}p^J_R)\ .
\eeq
%%%%%%%%%%%%%%
Thus, in the language of Narain torus compactification the
consistency condition for constructing orbifold models is that
$g$ must be an automorphism of the lattice
$\Gamma^{D,D}$ on which $(p^I_L,p^I_R)$ lies, i.e.
%%%%%%%%%%%%%% 3.18
\beq
g\ :\quad \Gamma^{D,D}\ \rightarrow \ \Gamma^{D,D}\ .
\eeq
%%%%%%%%%%%%%%%
As we have seen above, this condition is equivalent to the condition
(\ref{consist1}) together with the condition (\ref{consist2}).

Before closing this section, it may be instructive to present
some examples of $B^{IJ}$ and $U^{IJ}$, which satisfy the
conditions (\ref{consist1}) and (\ref{consist2}).
Let us consider the following $(D+D)$-dimensional Lorentzian even
self-dual lattice, which has been introduced by Englert and Neveu
\cite{Englert-N},
%%%%%%%%%%%%%%%%% 3.19
\beq
\Gamma^{D,D}\ =\ \{\ (p^I_L,p^I_R)\ |\ p^I_L,p^I_R\ \in\ \Lambda_W
({\cal G}),\ p^I_L-p^I_R\ \in\ \Lambda_R({\cal G})\ \}\ ,
\label{ENlattice}
\eeq
%%%%%%%%%%%%%%
where $\Lambda_R(\cal G)$ $(\Lambda_W(\cal G))$ is the root (weight)
lattice of a simply-laced Lie algebra $\cal G$ with rank $D$ and the
squared length of the root vectors is normalized to two.
In this normalization, the weight lattice $\Lambda_W(\cal G)$ is
just the dual lattice of $\Lambda_R(\cal G)$.
The lattice (\ref{ENlattice}) can be obtained through the relation
(\ref{relation1}) (or (\ref{relation2})) by choosing $\Lambda$
and $B^{IJ}$ as follows:
%%%%%%%%%%%%%%%%
\beq
\Lambda\ =\ \Lambda_R(\cal G)\ ,
\eeq
%%%%%%%%%%%%%%%%
and
%%%%%%%%%%%%%%
\beq
\alpha^I_i B^{IJ} \alpha^J_j\ =\ \alpha^I_i\alpha^I_j\quad
\mbox{mod}\ 2\ ,
\eeq
%%%%%%%%%%%%%%%%%
where $\alpha_i$ is a simple root of $\cal G$ which is normalized
to $(\alpha_i)^2=2$.
If we choose the rotation matrix $U$ to be an automorphism of the
root lattice $\Lambda_R(\cal G)$, i.e.
%%%%%%%%%%%%%%%%% 3.22
\beq
U^{IJ}w^J\ \in\ \Lambda_R({\cal G})\quad
\mbox{for all}\ \  w^I \in \Lambda_R({\cal G}),
\eeq
%%%%%%%%%%%%%%%%%
then the rotation $U$ is also an automorphism of the lattice
(\ref{ENlattice}) and satisfies the condition (\ref{consist2}), i.e.
%%%%%%%%%%%%%% 3.23
\beq
(B-U^TBU)^{IJ}w^J\ \in\ 2\Lambda_W({\cal G})\quad
\mbox{for all}\ \  w^I \in \Lambda_R({\cal G}).
\eeq
%%%%%%%%%%%%%%
The lattice (\ref{ENlattice}) is relevant to physics because gauge
symmetries may appear in the spectrum of closed string theory
through the Frenkel-Ka\v c-Segal mechanism
\cite{Frenkel-KS}, where
the lattice (\ref{ENlattice}) plays a crucial role.
Hence, our work would be of importance for the construction
of realistic orbifold models.

\vspace{1cm}

%%%%%%%%%%%%%%%%%%%%%%%%%%%%%%%%%%%%%%%%%%%%%%%%%%%%%%%%%%%%%%%%%%%%%%
%%%%%%%%%%%%%%%%%%%%%%%%%%%%%%%%%%%%%%%%%%%%%%%%%%%%%%%%%%%%%%%%%%%%%%

\sect{Quantization of Strings on Orbifolds}

In this section, we shall discuss the cocycle property of vertex
operators and show that zero modes of strings should obey nontrivial
quantization conditions.
We will also discuss the action of twist
operators on the string coordinate and find that
the left- and right-moving
coordinates transform anomalously for topologically nontrivial twists.

\subsection{The Cocycle Property of Vertex Operators and
            Quantization of Zero Modes}

Since on the orbifold a point \{$X^I\}$ is identified with
$\{g\cdot X^I+\pi w^I\}$ for all $w^I\in\Lambda$ and $g\in G$,
the string coordinate in general obeys the following boundary
condition:
%%%%%%%%%%%%%% 4.1
\beq
X^I(\tau,\sigma+\pi)\ =\ U^{IJ}X^J(\tau,\sigma)\ +\ \pi w^I\ .
\label{twistbc}
\eeq
%%%%%%%%%%%%%%
Strings which obey the boundary condition with $U^{IJ}=\delta^{IJ}$
are called untwisted strings. Strings with $U^{IJ}\ne \delta^{IJ}$
are called twisted strings.
In the following we will treat untwisted and twisted strings
in the same way.
The quantization of untwisted strings will be given by simply
putting $U^{IJ}=\delta^{IJ}$.

It is convenient to take the following coordinate system:
%%%%%%%%%%%%%%% 4.2
\beqs
U^{IJ} &=&
          \left(
           \begin{array}{cc}
            \delta^{ab} & 0 \\
            0 & \widetilde{U}^{\alpha\beta}
           \end{array}
           \right)^{IJ}\ ,\nonumber\\
X^I &=& (X^a,X^\alpha)\ ,
\label{cosystem}
\eeqs
%%%%%%%%%%%%%%%%%
where $I,J=1,\cdots,D,\  a,b=1,\cdots,d$
and $\alpha,\beta=1,\cdots,D-d.$
The matrix $\widetilde{U}^{\alpha\beta}$
is assumed to have no eigenvalues of one.
Then the mode expansion of $X^I(\tau,\sigma)$ will be given by
%%%%%%%%%%%%%%
\beqs
X^a(\tau,\sigma)\ &=&\ x^a + (p^a - B^{aJ}w^J)\tau + w^a\sigma
                   + X^a_{osc}(\tau,\sigma)\ ,  \nonumber\\
X^\alpha(\tau,\sigma)\ &=&\ x^\alpha + X^\alpha_{osc}(\tau,\sigma)\ ,
\eeqs
%%%%%%%%%%%%%%% 4.3
where $x^\alpha$ satisfies the fixed point equation,
%%%%%%%%%%%%%%  4.4
\beq
x^\alpha\ =\ \widetilde{U}^{\alpha\beta}x^\beta\ +\ \pi w^\alpha\ .
\label{fixedpoint}
\eeq
%%%%%%%%%%%%%%
The $X^I_{osc}(\tau,\sigma)$ denotes the oscillator part of the string
coordinate and satisfies
%%%%%%%%%%%%% 4.4
\beq
X^I_{osc}(\tau,\sigma + \pi)\ =\ U^{IJ}X^J_{osc}(\tau,\sigma)\ .
\eeq
%%%%%%%%%%%%%
Let $N$ be the smallest positive integer such that $g^N = 1$
and hence $U^N = {\bf 1}$.
The orthogonal matrix $U$ can be diagonalized  by a unitary matrix
$M$:
%%%%%%%%%%%%%% 4.5
\beq
M^{\dag}UM\ =\ U_{diag}\ ,\quad \mbox{with}\ M^{\dag}M = {\bf 1}\ ,
\eeq
%%%%%%%%%%%%%%
where
%%%%%%%%%%%%% 4.6
\beq
U^{IJ}_{diag}\ =\ \omega^{t_I}\delta^{IJ}\ ,
\quad (\omega = e^{i2\pi /N})\ .
\eeq
%%%%%%%%%%%%%
Here, we will choose $t_I$ such that $0 \le t_I \le N-1$.
Since $U$ is a real matrix, we have
%%%%%%%%%%%% 4.7
\beq
M^{T}UM^{*}\ =\ U^{*}_{diag}\ .
\eeq
%%%%%%%%%%%%
This implies that the set of $\{ \omega^{t_I}, I=1,\cdots,D\}$
is equivalent to the set of $\{ \omega^{-t_I}, I=1,\cdots,D\}$.
In terms of the eigenvalues of  $U$, $X^I_{osc}(\tau,\sigma)$
can be expanded as
%%%%%%%%%%%% 4.8
\beqs
X^I_{osc}(\tau,\sigma)&=&
   \frac i{2} \sum_{n_J\in {\bf Z}-\frac{t_J}{N}>0}
                \frac 1{n_J} \{M^{IJ}\gamma^J_{n_J}
                e^{-2in_J(\tau+\sigma)}
                            - M^{*IJ}\gamma^{J\dag}_{n_J}
                e^{2in_J(\tau+\sigma)}\} \nonumber\\
   &+&\frac i{2} \sum_{n_J\in {\bf Z}+\frac{t_J}{N}>0}
                \frac 1{n_J} \{M^{IJ}\overline{\gamma}^J_{n_J}
                e^{-2in_J(\tau-\sigma)}
                            - M^{*IJ}\overline{\gamma}^{J\dag}_{n_J}
                e^{2in_J(\tau-\sigma)}\}\ .
\eeqs
%%%%%%%%%%%%%
The quantization conditions of the oscillators are given as usual,
%%%%%%%%%%% 4.9
\beqs
[\ \gamma^I_{m_I}\ ,\ \gamma^{J\dag}_{n_J}\ ]\ &=&
         m_I\delta^{IJ}\delta_{m_I,n_J}\ ,
         \quad\mbox{for}\ m_I\in {\bf Z}-\frac{t_I}{N}>0\
         \mbox{and}\ n_J\in {\bf Z}-\frac{t_J}{N}>0\ ,\nonumber\\
{[\ \overline{\gamma}^I_{m_I}\ ,
         \ \overline{\gamma}^{J\dag}_{n_J}\ ]}\ &=&
         m_I\delta^{IJ}\delta_{m_I,n_J}\ ,
         \quad\mbox{for}\ m_I\in {\bf Z}+\frac{t_I}{N}>0\
         \mbox{and}\ n_J\in {\bf Z}+\frac{t_J}{N}>0\ , \nonumber\\
\mbox{otherwise} & & \mbox{zeros}\ .
\eeqs
%%%%%%%%%%%%%
In the following, we will discuss the cocycle property of
vertex operators in detail and show that the zero modes of
strings obey nontrivial quantization conditions.

Let us consider a vertex operator $V(k_L,k_R;z)$ which describes
the emission of an untwisted state with the momentum
$(k^I_L,k^I_R) \in \Gamma^{D,D}.$  The vertex operator will be
of the form (see appendix A for the precise definition),
%%%%%%%%%%%%%% 4.10
\beq
V(k_L,k_R;z) = f_{k_L,k_R}(z)\ :\exp\{ik_L^IX_L^I(z)
                + ik_R^IX_R^I(\overline{z})\}:\ ,
\eeq
%%%%%%%%%%%%%%
where $X^I_L(z)$ and $X^I_R(\overline{z})$ are the left- and right-moving
string coordinates defined by
%%%%%%%%%%%%
\beqs
X^I_L(z) &=& x^I_L - ip^I_L\ln z +
             i \sum_{n_J\in {\bf Z}-\frac{t_J}{N}>0}
                \frac 1{n_J} \{M^{IJ}\gamma^J_{n_J}z^{-n_J}
                 - M^{*IJ}\gamma^{J\dag}_{n_J}z^{n_J}\}\ , \nonumber\\
X^I_R(\overline{z}) &=& x^I_R - ip^I_R\ln \overline{z} +
             i \sum_{n_J\in {\bf Z}+\frac{t_J}{N}>0}
                \frac 1{n_J} \{M^{IJ}\overline{\gamma}^J_{n_J}
                \overline{z}^{-n_J}
                 - M^{*IJ}\overline{\gamma}^{J\dag}_{n_J}
                \overline{z}^{n_J}\}\ ,
\eeqs
%%%%%%%%%%%%
with $p^\alpha_L=p^\alpha_R=0\ (\alpha=1,\cdots,D-d).$
The relations between $x^I_L,x^I_R,p^a_L,p^a_R$ and
$x^I,Q^I,p^a,w^I$ are given by
%%%%%%%%%%%%
\begin{eqnarray}
x^I_L&=&(1-B)^{IJ}x^J+Q^I\ ,\nonumber\\
x^I_R&=&(1+B)^{IJ}x^J-Q^I\ ,\nonumber\\
p^a_L&=&\frac1{2}(p^a-B^{aJ}w^J)+\frac1{2}w^a\ ,\nonumber\\
p^a_R&=&\frac1{2}(p^a-B^{aJ}w^J)-\frac1{2}w^a\ .
\end{eqnarray}
%%%%%%%%%%%%%%%%%%
The $f_{k_L,k_R}(z)$ is a normalization factor and in general
depends on $k^I_L,k^I_R$ and $z$ in twisted sectors.
The product of two vertex operators
%%%%%%%%%%%% 4.11
\beq
V(k'_L,k'_R;z')V(k_L,k_R;z)\ ,
\label{VV}
\eeq
%%%%%%%%%%%%%%
is well-defined if $|z'| > |z|.$
The different ordering of the two vertex operators corresponds
to the different \lq\lq time"-ordering.
To obtain scattering amplitudes, we must sum over all possible
\lq\lq time"-orderings for the emission of states.
We must then establish that each contribution is independent
of the order of the vertex operators to enlarge the region of
integrations over $z$ variables
\cite{Schwarz-S}.
Thus the product (\ref{VV}), with respect to $z$ and $z'$,
has to be analytically continued to the region $|z|>|z'|$
and to be identical to
%%%%%%%%%%%% 4.12
\beq
V(k_L,k_R;z)V(k'_L,k'_R;z')\ ,
\eeq
%%%%%%%%%%%%%%
for $|z|>|z'|.$
In terms of the zero modes, the above statement can be translated into
the following condition:
%%%%%%%%%%%%%% 4.13
\beq
V_0(k'_L,k'_R)V_0(k_L,k_R) = \eta\: V_0(k_L,k_R)V_0(k'_L,k'_R)\ ,
\label{cocycle}
\eeq
%%%%%%%%%%%%%
where
%%%%%%%%%%%%%% 4.14
\beq
V_0(k_L,k_R) = \exp\{ik_L^I\widehat{x}_L^I +
              ik_R^I\widehat{x}_R^I \}\ .
\eeq
%%%%%%%%%%%%
The wedge $\wedge$ may be attached to operators to distinguish
$q$-numbers from $c$-numbers.
The phase $\eta$ is given by
\cite{Kac-P}
%%%%%%%%%%%%%%5 4.15
\beq
\eta = \exp\Bigl\{ i\pi k'^I_L\Bigl( 1 + \sum^{N}_{m = 1}
       \frac{m}{N}(U^{-m} - U^{m})\Bigr)^{IJ}k^J_L
            - i\pi k'^I_R\Bigl( 1 + \sum^{N}_{m = 1}
       \frac{m}{N}(U^{-m} - U^{m})\Bigr)^{IJ}k^J_R
       \Bigr\}\ .
\label{cophase}
\eeq
%%%%%%%%%%%%%%%
Since the relation (\ref{cocycle}) is a key equation in our
discussions,
we will explicitly derive eq. (\ref{cocycle})
with eq. (\ref{cophase}) in appendix A.
In the untwisted sector $(U^{IJ}=\delta^{IJ})$, $\eta$ reduces
to the well-known result
\cite{Frenkel-KS,Goddard-O},
%%%%%%%%%%%% 4.16
\beq
\eta = (-1)^{k'_L\cdot k_L-k'_R\cdot k_R}\ .
\eeq
%%%%%%%%%%%%
To satisfy the relation (\ref{cocycle}), one might attach a
cocycle operator to the vertex operator.
However, the introduction of a cocycle operator might cause troubles
with the construction of the Hilbert space since we need an explicit
realization of the cocycle operator to construct the Hilbert space.
Here, instead of introducing cocycle operators, we will follow the
approach given by ref.
\cite{Sakamoto-IKK,Sakamoto89}.

As mentioned in sect. 1, we have no definite prescription
how to quantize the \lq\lq coordinate" $Q^I$ due to the lack
of geometrical meanings of $Q^I$.
Here, we will determine the quantization of $Q^I$ by requiring
the relation (\ref{cocycle}).
It turns out that the following commutation relations satisfy the
relation (\ref{cocycle}) with the correct phase (\ref{cophase}):
\footnote{
 One might assume, instead of eqs. (\ref{quant1}), the following
commutation relations:
%%%%%%
\beqs
 [\ \widehat{x}^I_L\ ,\ \widehat{x}^J_L\ ] &=&
         -i\pi \Bigl( 1 + \sum^{N}_{m =1}\frac{m}{N}
         (U^{-m} - U^{m})\Bigr)^{IJ}\ ,\nonumber\\
 {[\ \widehat{x}^I_R\ ,\ \widehat{x}^J_R\ ]} &=&
         i\pi \Bigl(1 + \sum^{N}_{m =1}\frac{m}{N}
         (U^{-m} - U^{m})\Bigr)^{IJ}\ ,\nonumber\\
 {[\ \widehat{x}^I_L\ ,\ \widehat{x}^J_R\ ]} &=&
         0\ .\nonumber
\eeqs
%%%%%%%%%%%%%%%%%
These are not, however,  allowed because the first two
commutation relations
must be antisymmetric with respect to the superscripts $I$ and $J.$
}
%%%%%%%%%%%%%%%% 4.17
\beqs
 [\ \widehat{x}^I_L\ ,\ \widehat{x}^J_L\ ] &=&
         i\pi \Bigl(B - \sum^{N}_{m =1}\frac{m}{N}
         (U^{-m} - U^{m})\Bigr)^{IJ}\ ,\nonumber\\
 {[\ \widehat{x}^I_R\ ,\ \widehat{x}^J_R\ ]} &=&
         i\pi \Bigl(B + \sum^{N}_{m =1}\frac{m}{N}
         (U^{-m} - U^{m})\Bigr)^{IJ}\ ,\nonumber\\
 {[\ \widehat{x}^I_L\ ,\ \widehat{x}^J_R\ ]} &=&
         i\pi (1 - B)^{IJ}\ .
\label{quant1}
\eeqs
%%%%%%%%%%%%%%%%%%5
To see this, we will use the fact that
%%%%%%%%%%%%%%% 4.18
\beqs
& &(1+B)^{IJ}k^J_L + (1-B)^{IJ}k^J_R \ \in \ 2\Lambda^{*}\ ,
\nonumber\\
& &k^I_L - k^I_R\ \in\ \Lambda\ .
\eeqs
%%%%%%%%%%%%%%
To discuss physical or geometrical meanings of the relations
(\ref{quant1}), let us rewrite eqs. (\ref{quant1}), in terms of $x^I$
and $Q^I$, as
%%%%%%%%%%%%%% 4.19
\beqs
 [\ \widehat{x}^I\ ,\ \widehat{x}^J\ ] &=& 0\ ,\nonumber\\
 {[\ \widehat{x}^I\ ,\ \widehat{Q}^J\ ]} &=&
   -i\frac{\pi}{2}\Bigl( 1 +  \sum^{N}_{m =1}\frac{m}{N}
         (U^{-m} - U^{m})\Bigr)^{IJ}\ ,\nonumber\\
 {[\ \widehat{Q}^I\ ,\ \widehat{Q}^J\ ]} &=&
     i\frac{\pi}{2} [\ \sum^{N}_{m =1}\frac{m}{N}
         (U^{-m} - U^{m})\ ,\ B\ ]^{IJ}\ .
\label{quant2}
\eeqs
%%%%%%%%%%%%%%
For completeness, we list other nonzero commutation relations of the
zero modes,
%%%%%%%%%%%%% 4.20
\beqs
[\ \widehat{x}^a\ ,\ \widehat{p}^b\ ] &=& i\delta^{ab}\ ,\nonumber\\
{[\ \widehat{Q}^I\ ,\ \widehat{w}^J\ ]} &=& i\delta^{IJ}\ .
\label{quant3}
\eeqs
%%%%%%%%%%%%%
The relations (\ref{quant2}) and (\ref{quant3}) agree with the results
of ref.
\cite{Sakamoto89} for $U^{IJ}=\delta^{IJ}$ and those of ref.
\cite{Sakamoto-IKK} for
$[U,B]=0$ and $\det(1-U)\ne 0$.
In the untwisted sector $(U^{IJ}=\delta^{IJ}),$ the relations
(\ref{quant2}) reduce to
%%%%%%%%%%% 4.21
\beqs
[\ \widehat{x}^I\ ,\ \widehat{x}^J\ ] &=& 0\ ,\nonumber\\
{[\ \widehat{x}^I\ ,\ \widehat{Q}^J\ ]}
&=& -i\frac{\pi}{2}\delta^{IJ}\ ,\nonumber\\
{[\ \widehat{Q}^I\ ,\ \widehat{Q}^J\ ]} &=& 0\ .
\label{quant4}
\eeqs
%%%%%%%%%%%%
The second relation of eqs. (\ref{quant4}) seems strange but it has
a simple geometrical interpretation
\cite{Sakamoto89}:
Let $|x^I,w^I=0>$ be a string state whose center of mass coordinate
and winding number are given by $x^I$ and $w^I=0$, respectively.
The operator $\exp\{iw\cdot \widehat{Q}\}$ is a shift operator of the
winding number by $w^I.$
It follows from the second equation of (\ref{quant4}) that
%%%%%%%%%%%% 4.22
\beq
[\ \widehat{x}^I\ ,\ \exp\{iw\cdot \widehat{Q}\}\ ] =
\frac{\pi}{2}w^I \exp\{iw\cdot \widehat{Q}\}\ ,
\eeq
%%%%%%%%%%%%%%
which implies that the state $\exp\{iw\cdot \widehat{Q}\} |x^I,w^I=0>$
has the eigenvalues of $x^I+(\pi /2)w^I$ and $w^I$ for the
center of mass coordinate and the winding number, respectively, i.e.
%%%%%%%%%%%% 4.23
\beq
\exp\{iw\cdot \widehat{Q}\} |x^I,w^I=0>\
     \propto\ |x^I+\frac{\pi}{2}w^I,w^I>\ .
\eeq
%%%%%%%%%%%%
At first sight it seems strange that the center of mass coordinate
of a string changes by $(\pi /2)w^I$ when the winding number changes
by $w^I.$
This is, however, acceptable geometrically.
We first note that the \lq\lq center of mass coordinate" of a string
{\it on the torus\/} is {\it not\/} a well-defined notion in the
presence of nonzero winding numbers because the string winds
around the torus.
However, it will still be a useful notion {\it on the covering space
of the torus\/} and it may be natural to say that the
\lq\lq center of mass coordinate" of the string is located
at $x^I+(\pi /2)w^I$ rather than $x^I$ in the presence of the
winding number if we define the \lq\lq center of mass coordinate" by
%%%%%%%%%%%%%%% 4.24
\beq
\int^{\pi}_{0} \frac{d\sigma}{\pi} X^I(\tau=0,\sigma)
= x^I + \frac{\pi}{2}w^I\ .
\eeq
%%%%%%%%%%%%%%
In the twisted sector ($U^{IJ}\ne \delta^{IJ}$), the second relation
of eqs. (\ref{quant2}) also has a simple interpretation.
In the coordinate system (\ref{cosystem}), the second relation of eqs.
(\ref{quant2}) can be rewritten as
%%%%%%%%%% 4.25
\beqs
[\ \widehat{x}^a\ ,\ \widehat{Q}^b\ ]
  &=& -i\frac{\pi}{2}\delta^{ab}\ ,\nonumber\\
{[\ \widehat{x}^\alpha\ ,\ \widehat{Q}^\beta\ ]}
   &=& -i\pi \Bigl(\frac{1}{1-\widetilde{U}}\Bigr)^{\alpha\beta}\ ,
    \nonumber\\
\mbox{otherwise} & & \mbox{zeros}\ ,
\label{quant5}
\eeqs
%%%%%%%%%%%%%%
where we have used the identities
%%%%%%%%%% 4.26
\beqs
\Bigl(\frac{1}{1-\widetilde{U}}\Bigr)^{\alpha\beta}
&=& \sum^{N}_{m =1} \frac{m}{N} (\widetilde{U}^{-m})^{\alpha\beta}\ ,
\nonumber\\
\sum^{N}_{m=1} (\widetilde{U}^{m})^{\alpha\beta} &=& 0\ .
\eeqs
%%%%%%%%%%%%
The first relation of eqs. (\ref{quant5}) can be understood in the
same way as the untwisted sector.
The second relation of eqs. (\ref{quant5}) is consistent with the
constraint (\ref{fixedpoint}).
Thus, the commutation relations of the zero modes have
simple geometrical meanings except for the last relation of eqs.
(\ref{quant2}).
We could not find any simple explanation for the last commutation
relation of eqs. (\ref{quant2}) due to the lack of geometrical
meanings of $Q^I$.
We will, however, show in sect. 6 that
all the commutation relations of the zero modes derived
above are consistent with modular invariance of partition
functions.

%%%%%%%%%%%%%%%%%%%%%%%%%%%%%%%%%%%%%%%%%%%%%%%%%%%%%%%%%%%%%%%%%%%%

\subsection{Anomalous Transformations}

Let $h$ be an element of $G$ associated with an orthogonal matrix $V$.
To calculate one loop partition functions in the $g$-sector, we
need only the elements of $G$ which commute with $g$.
Thus we will restrict our considerations to the elements $h$ of $G$
such that
%%%%%%%%%%%% 4.27
\beq
hg = gh\ .
\eeq
%%%%%%%%%%%%
This implies that the orthogonal matrix $V$ commutes with $U$, i.e.
%%%%%%%%%%%% 4.28
\beq
VU = UV\ .
\label{VU=UV}
\eeq
%%%%%%%%%%%%%%%
In the coordinate system (\ref{cosystem}), the matrix $V^{IJ}$
will be of the form,
%%%%%%%%
\beq
V^{IJ} =
          \left(
           \begin{array}{cc}
            \widetilde{V}^{ab} & 0 \\
            0 & \widetilde{V}^{\alpha\beta}
           \end{array}
           \right)^{IJ}\ .
\eeq
%%%%%%%%%%%%%%%%%
The action of $h$ on $X^I(\tau,\sigma)$ in the $g$-sector is given by
%%%%%%%%%%%% 4.29
\beq
h:\quad X^I(\tau,\sigma)\ \rightarrow\ V^{IJ}X^J(\tau,\sigma)\ .
\label{htransf}
\eeq
%%%%%%%%%%%%%%%%%
Note that the above transformation is consistent with the boundary
condition (\ref{twistbc}) only if eq. (\ref{VU=UV}) is satisfied.
In terms of the zero modes, the transformation (\ref{htransf})
can be written as
%%%%%%%%%%%%%% 4.30
\beqs
h:\quad x^I &\rightarrow& V^{IJ}x^J\ ,\nonumber\\
        p^a &\rightarrow& \widetilde{V}^{ab}p^b
            - [V,B]^{aJ}w^J\ ,\nonumber\\
        w^I &\rightarrow& V^{IJ}w^J\ .
\label{normaltr}
\eeqs
%%%%%%%%%%%%%%%%%
It should be emphasized that eq. (\ref{htransf}) does not
completely specify the
action of $h$ on $Q^I$ since $Q^I$ does not appear in the mode
expansion of $X^I(\tau,\sigma)$.
Our purpose in this subsection is to determine the action of
$h$ on $Q^I$.

The relevant operators in string theory are the momentum operators,
$P^I_L(z)\equiv i\partial_zX^I_L(z)$ and
$P^I_R(\overline{z})\equiv
i\partial_{\overline{z}}X^I_R(\overline{z})$, and the
vertex operator $V(k_L,k_R;z)$.
Other operators can be obtained from the operator products of these
operators.
For example the energy-momentum tensors of the left- and right-movers
are given by
%%%%%%%%%% 4.31
\beqs
T(z) &=& \lim_{w\to z}\frac{1}{2}\Bigl(
       P^I_L(w)P^I_L(z) - \frac{D}{(w-z)^2}\Bigr)\ ,\nonumber\\
\overline{T}(\overline{z})
     &=& \lim_{\overline{w}\to \overline{z}}\frac{1}{2}\Bigl(
       P^I_R(\overline{w})P^I_R(\overline{z}) -
       \frac{D}{(\overline{w}-\overline{z})^2}\Bigr)\ .
\eeqs
%%%%%%%%%%%%%%
Under the action of $h$, the momentum operators transform as
%%%%%%%%%% 4.32
\beq
h :\quad (P^I_L(z) , P^I_R(\overline{z}) )\  \rightarrow\
         (V^{IJ}P^J_L(z) , V^{IJ}P^J_R(\overline{z}) )\ ,
\eeq
%%%%%%%%%%%%%%%
which leaves the energy-momentum tensors invariant, as it should be.
One might expect that under the action of $h$ the left- and
right-moving string coordinates, $X^I_L(z)$ and $X^I_R(\overline{z})$,
would transform in the same way as $P^I_L(z)$
and $P^I_R(\overline{z})$.
However, it is not the case if $V^{IJ}$ does not commute with $B^{IJ}.$
Suppose that under the action of $h$, $\widehat{x}^I_L$
and $\widehat{x}^I_R$ would
transform as
%%%%%%%% 4.33
\beqs
h :\quad \widehat{x}^I_L\ &\rightarrow&\ V^{IJ}\widehat{x}^J_L\ ,
          \nonumber\\
         \widehat{x}^I_R\ &\rightarrow&\ V^{IJ}\widehat{x}^J_R\ .
\label{xtoVx}
\eeqs
%%%%%%%%%%%%%
It is easy to see that the transformations (\ref{xtoVx}) is
inconsistent with the commutation relations (\ref{quant1}) unless
$V^{IJ}$ commutes with $B^{IJ}$.
In ref.
\cite{Sakamoto-TH}, it has been shown that in the untwisted sector the
correct action of $h$ on $\widehat{x}^I_L$ and
$\widehat{x}^I_R$ is given by
\footnote{Set $U_L=U_R\equiv V$ and assume the form (6.13) in the
first reference of ref.
\cite{Sakamoto-TH}.}
%%%%%%%%% 4.34
\beqs
h :\quad \widehat{x}^I_L\ &\rightarrow& V^{IJ}\widehat{x}^J_L
         + \pi V^{IJ}\Bigl(\frac{1}{2}(B-V^TBV)-C_h\Bigr)^{JK}
         \widehat{w}^K\ ,\nonumber\\
         \widehat{x}^I_R\ &\rightarrow& V^{IJ}\widehat{x}^J_R
         - \pi V^{IJ}\Bigl(\frac{1}{2}(B-V^TBV)-C_h\Bigr)^{JK}
         \widehat{w}^K\ ,
\label{anomaltr}
\eeqs
%%%%%%%%%%%%%%
where $C^{IJ}_h$ is a symmetric matrix defined through the relation
%%%%%%%%% 4.35
\beq
w^IC^{IJ}_hw'^J = \frac{1}{2}w^I(B-V^TBV)^{IJ}w'^J\quad
    \mbox{mod}\ 2\ ,
\label{Ch}
\eeq
%%%%%%%%%%%
for all $w^I,w'^I \in \Lambda.$
The existence of a symmetric matrix $C^{IJ}_h$ is guaranteed by the
fact that
%%%%%%%%%% 4.36
\beq
\frac{1}{2}w^I(B-V^TBV)^{IJ}w'^J \in {\bf Z}\quad
  \mbox{for all}\ w^I,w'^I \in \Lambda\ .
\eeq
%%%%%%%%%%%%%
Note that the anomalous terms in eqs. (\ref{anomaltr}) may
disappear if $V^{IJ}$ commutes with $B^{IJ}$.
In ref.
\cite{Sakamoto-TH}, the transformation (\ref{anomaltr}) in the untwisted
sector has been derived from the requirement that the action of
$h$ on vertex operators has to be consistent with the cocycle
property of vertex operators.
The transformation (\ref{anomaltr}) is also correct in the
$g$-twisted sector:
It is easy to see that the transformations (\ref{anomaltr}) and
(\ref{normaltr}) are consistent with the commutation relations
(\ref{quant2}) and (\ref{quant3}) and also with the relation
(\ref{cocycle}).

It seems that the transformation (\ref{anomaltr}) loses its
geometrical meaning unless $V^{IJ}$ commutes with $B^{IJ}$.
However, the transformation (\ref{anomaltr}) is still
geometrically well-defined because the anomalous terms in eq.
(\ref{anomaltr}) can be regarded as a torus shift, i.e.
%%%%%%%%% 4.37
\beq
h :\quad (\widehat{x}^I_L,\widehat{x}^I_R)\ \rightarrow\
         (V^{IJ}\widehat{x}^J_L,V^{IJ}\widehat{x}^J_R)
         + \mbox{torus shift}\ .
\eeq
%%%%%%%%%%%%
To see this, let us consider the action of $h$ on
$k_L\cdot\widehat{x}_L + k_R\cdot\widehat{x}_R$
for $(k^I_L,k^I_R) \in
\Gamma^{D,D}.$
%%%%%%%%%% 4.38
\beq
h : k_L\cdot\widehat{x}_L + k_R\cdot\widehat{x}_R\ \rightarrow\
   k_L\cdot V\widehat{x}_L + k_R\cdot V\widehat{x}_R
   + \pi (k_L-k_R)\cdot V(\frac{1}{2}(B-V^TBV)-C_h)\widehat{w}\ .
\eeq
%%%%%%%%%%%%%%%
The last term of the right hand side may be regarded as a trivial
shift because
%%%%%%%%%% 4.39
\beq
(k_L-k_R)\cdot V(\frac{1}{2}(B-V^TBV)-C_h)\widehat{w} = 0
\quad \mbox{mod}\ 2\ ,
\label{trshift}
\eeq
%%%%%%%%%%%%%%%
for all $(k^I_L,k^I_R) \in \Gamma^{D,D}$ and $\widehat{w}^I \in \Lambda.$
(Recall that $k^I_L-k^I_R \in \Lambda.$)
It may be instructive to repeat the above discussion in terms of
$\widehat{x}^I$ and $\widehat{Q}^I$.
The transformation (\ref{anomaltr}) can be rewritten as
%%%%%%%%%% 4.40
\beqs
h :\quad {\widehat x}^I &\rightarrow& V^{IJ}{\widehat x}^J\ ,
            \nonumber\\
         {\widehat Q}^I &\rightarrow& V^{IJ}{\widehat Q}^J -
         [V,B]^{IJ}{\widehat x}^J
         + \pi V^{IJ} \Bigl(\frac{1}{2}(B-V^TBV)-C_h
         \Bigr)^{JK}{\widehat w}^K\ .
\label{anomaltr2}
\eeqs
%%%%%%%%%%%%%%%
Note that the anomalous term appears only in the transformation of
$Q^I$.
The anomalous term may be regarded as a trivial shift because
%%%%%%%%%% 4.41
\beq
V^{IJ}\Bigl(\frac{1}{2}(B-V^TBV)-C_h
      \Bigr)^{JK}\widehat{w}^K\ \in\ 2\Lambda^{*}\ ,
\eeq
%%%%%%%%%%%%%%%
and because $Q^I$ may be regarded as a coordinate on a torus with the
identification of $Q^I \sim Q^I+\pi p^I$ for all $p^I \in 2\Lambda^{*}.$
(Recall that $Q^I$ is a conjugate variable to $w^I$ and that
$w^I \in \Lambda.$)

What roles do the anomalous terms in eq. (\ref{anomaltr}) or
(\ref{anomaltr2}) play?
To answer this question, let us consider the action of $h$ on the
vertex operator.
The result is
%%%%%%%%%%%% 4.42
\beq
h :\quad V(k_L,k_R;z)\ \rightarrow\ \rho\: V(V^Tk_L,V^Tk_R;z)\ ,
\label{hV}
\eeq
%%%%%%%%%%%%%%%%%
where
%%%%%%%%% 4.43
\beq
\rho = \exp\Bigl\{-i\frac{\pi}{2}(k_L-k_R)\cdot VC_hV^T(k_L-k_R)
           \Bigr\}\ ,
\label{Z2phase}
\eeq
%%%%%%%%%%%%%%%
with $\rho^2 = 1.$
It should be noted that we cannot
use the relation (\ref{trshift}) directly in the exponent to derive
the relation (\ref{hV}) because
$\widehat{w}^I$ is not a $c$-number but a $q$-number.
We can use the relation (\ref{trshift}) only after separating the
terms depending on $\widehat{w}^I$ from $V(V^Tk_L,V^Tk_R;z)$ by use
of the Hausdorff formula.
The phase factor $\rho$ appearing in the transformation (\ref{hV})
plays an important role in extracting physical states.
To see this, let us consider a state
$|k_L,k_R> \equiv V(k_L,k_R;z=0)|A>$, where $|A>$ is an arbitrary
$h$-invariant state.
Suppose that the momentum $(k^I_L,k^I_R)$ is $h$-invariant, i.e.
$(k^I_L,k^I_R) = (V^{IJ}k^J_L,V^{IJ}k^J_R).$
Then we find
%%%%%%%%%%% 4.44
\beq
h :\quad |k_L,k_R>\ \rightarrow\ \rho\: |k_L,k_R>\ .
\eeq
%%%%%%%%%%%%%
On the orbifold, any physical states must be $h$-invariant for all
$h \in G.$
Therefore, the state $|k_L,k_R>$ must be removed from the physical
Hilbert space unless $\rho = 1.$
It may be worthwhile noting
that in an algebraic point of view the phase
$\rho$ in eq. (\ref{hV}) has a connection with automorphisms of
algebras rather than automorphisms of lattices.

Before closing this subsection, we should make a comment on the
transformations (\ref{anomaltr}) or (\ref{anomaltr2}).
We might add some extra terms,
which are consistent with eqs. (\ref{quant1}), to the transformations
(\ref{anomaltr}) or (\ref{anomaltr2}).
The extra terms would, however, make ill-defined the action of $h$
on the vertex operator and also would destroy modular invariance
of partition functions.

\vspace{1cm}

%%%%%%%%%%%%%%%%%%%%%%%%%%%%%%%%%%%%%%%%%%%%%%%%%%%%%%%%%%%%%%%%%%%%%%%%
%%%%%%%%%%%%%%%%%%%%%%%%%%%%%%%%%%%%%%%%%%%%%%%%%%%%%%%%%%%%%%%%%%%%%%%%

\sect{The Hilbert Space of the Zero Modes}

\subsection{Aharonov-Bohm like Effects}

We might observe some similarities between strings on orbifolds in the
presence of the antisymmetric background field and electrons
in the presence of an infinitely long solenoid.
Both underlying spaces are not simply-connected and possess
non-contractible loops.
The antisymmetric background field $B^{IJ}$
may play a similar role of
an external gauge field.
If an electron moves around the solenoid, a wave function of the
electron in general acquires a phase.
It may be natural to ask whether Aharonov-Bohm like effects occur
in our system.
We can show that a wave function $\Psi(x^I)$ of a twisted string
in the $g$-twisted sector is not, in general, periodic with respect to
a torus shift but
%%%%%%%% 5.1
\beq
\Psi(x^I+\pi w^I) = \exp\{-i\frac{\pi}{2}w^IC^{IJ}_gw^J\}\
                    \Psi(x^I)\ ,
\label{AB}
\eeq
%%%%%%%%%%%%%%%
where
%%%%%%%%% 5.2
\beq
w^I \in \Lambda_U = \{\ w^I \in \Lambda\ |\ w^I = U^{IJ}w^J\ \}\ .
\label{LambdaU}
\eeq
%%%%%%%%%%%%%%
The symmetric matrix $C^{IJ}_g$ is defined by
%%%%%%%% 5.3
\beq
w^IC^{IJ}_gw'^J = \frac{1}{2}w^I(B-U^TBU)^{IJ}w'^J \quad
\mbox{mod}\ 2\ ,
\label{Cg}
\eeq
%%%%%%%%%%%%%%%
for all $w^I,w'^I \in \Lambda.$
The phase in eq. (\ref{AB})
is equal to $\pm 1$ in general and will be equal to 1 if $[U,B] = 0.$
What are physical implications of the relation (\ref{AB})?
To see this, we first note that the left hand side of eq. (\ref{AB})
can be expressed as
%%%%%%%%% 5.4
\beq
\Psi(x^I+\pi w^I) = \exp\{-i\pi w^I\widehat{p}^I_{_{/\!/}}\}\
                    \Psi(x^I)\ ,
\label{AB2}
\eeq
%%%%%%%%%%%%%
where $\widehat{p}^I_{_{/\!/}}$ is the canonical momentum
restricted to the $U$-invariant subspace, i.e.
$\widehat{p}^I_{_{/\!/}} = (\widehat{p}^a,\widehat{p}^\alpha=0).$
Let us introduce a vector $s^I_g$ with $s^I_g = U^{IJ}s^J_g$
through the relation
%%%%%%%%%%% 5.5
\beq
w^Is^I_g = \frac{1}{2}w^IC^{IJ}_gw^J\quad \mbox{mod}\ 2\ ,
\label{sg}
\eeq
%%%%%%%%%%%%%
for all $w^I \in \Lambda_U.$
The existence of such a vector $s^I_g$ is guaranteed by the
following observation: Define
%%%%%%%%%% 5.6
\beq
f(w) \equiv \frac{1}{2}w^IC^{IJ}_gw^J\ .
\eeq
%%%%%%%%%%%%%
Then we find
%%%%%%%%% 5.7
\beqs
f(w+w') &=& f(w) + f(w') + w^IC^{IJ}_gw'^J \nonumber\\
        &=& f(w) + f(w')\qquad \mbox{mod}\ 2\quad
\mbox{for all}\ w^I,w'^I \in \Lambda_U\ ,
\eeqs
%%%%%%%%%%%%%%%
where we have used eqs. (\ref{LambdaU}) and (\ref{Cg}).
Comparing eq. (\ref{AB}) with eq. (\ref{AB2}), we conclude that
%%%%%%%%%% 5.8
\beq
\widehat{p}^I_{_{/\! /}} \in 2{\Lambda_U}^{*} + s^I_g\ ,
\label{momshift}
\eeq
%%%%%%%%%%%%%%%
where ${\Lambda_U}^{*}$ is the dual lattice of $\Lambda_U.$
Thus allowed eigenvalues of $\widehat{p}^I_{_{/\! /}}$ are different
from naively expected values
\cite{asymorb1} by $s^I_g.$
This result exactly agrees with the one expected from the
argument of modular invariance in ref.
\cite{Sakamoto-TH}.
In the next subsection we shall discuss eigenvalues of the zero
modes of the twisted strings in detail.

Before deriving eq. (\ref{AB}) or (\ref{momshift}),
it may be instructive to recall quantum mechanics on a circle
with a radius $L$, where a point $x$ is identified with $x+2\pi nL$
for all $n \in {\bf Z}.$
It is important to understand that the coordinate operator
$\widehat{x}$
itself is not well-defined on the circle while the canonical
momentum operator $\widehat p$ is well-defined.
However, the following operator:
%%%%%%%%% 5.9
\beq
\exp\{ik\widehat x\} \qquad \mbox{with}\ \ k = \frac{m}{L}\ \
(m \in {\bf Z})\ ,
\label{physop}
\eeq
%%%%%%%%%%%%%%
is well-defined on the circle because it is consistent with the torus
identification.
Physically the operator (\ref{physop}) plays a role of a momentum
shift by $k=m/L.$
It turns out that an operator $\exp\{i2\pi L\widehat p\}$ commutes
with \lq\lq all" operators, i.e. $\widehat p$ and
$\exp\{ik\widehat x\}$,
and hence it must be a $c$-number.
Indeed we have
%%%%%%%% 5.10
\beq
\exp\{i2\pi L\widehat p\} = 1\ .
\label{momquant}
\eeq
%%%%%%%%%%%%%%%
This leads to the well-known momentum quantization,
%%%%%%%% 5.11
\beq
\widehat p = \frac{m}{L} \qquad \mbox{with}\ \ m \in {\bf Z}\ .
\eeq
%%%%%%%%%%%%%%
In string theory on the orbifold, $\widehat p$  and
$\exp\{ik\widehat x\}$
will be replaced by the momentum operators $P^I_L(z),
P^I_R(\overline{z})$
and the vertex operator $V(k_L,k_R;z)$, respectively, and
an analog of the identity (\ref{momquant}) in the $g$-sector is
%%%%%%%%%% 5.12
\beqs
I_{k_L,k_R} &\equiv&
  \xi_{k_L,k_R} \exp\{-ik^I_LU^{IJ}\widehat{x}^J_L
    -ik^I_RU^{IJ}\widehat{x}^J_R\}
  \exp\{ik^I_L\widehat{x}^I_L + ik^I_R\widehat{x}^I_R\}   \nonumber\\
& &\quad \times
    \exp\{i2\pi(k^a_L\widehat{p}^a_L - k^a_R\widehat{p}^a_R)\}
    \nonumber\\
  &=& 1\ ,
\label{identity}
\eeqs
%%%%%%%%%%%%%%%%
where $(k^I_L,k^I_R) \in \Gamma^{D,D}$ and
%%%%%%%%%% 5.13
\beq
\xi_{k_L,k_R} = \exp\Bigl\{i\pi ((k^a_L)^2 - (k^a_R)^2)
    - i\frac{\pi}{2}(k_L-k_R)^I(UC_gU^T)^{IJ}(k_L-k_R)^J\Bigr\}\ .
\eeq
%%%%%%%%%%%%%%%
To prove the identity (\ref{identity}), it is important to understand
that the relevant operators in string theory on the orbifold are
$P^I_L(z), P^I_R(\overline{z})$ and $V(k_L,k_R;z)$ but not
$X^I_L(z), X^I_R(\overline{z})$ because
$P^I_L(z), P^I_R(\overline{z})$ and $V(k_L,k_R;z)$
are well-defined on the orbifold but $X^I_L(z), X^I_R(\overline{z})$
are not.
It is not difficult to show that $I_{k_L,k_R}$ defined in eq.
(\ref{identity}) commutes with \lq\lq all" operators, i.e.
$P^I_L(z), P^I_R(\overline{z})$ and $V(k_L,k_R;z)$ for all
$(k^I_L,k^I_R)\in \Gamma^{D,D}$,
by use of the commutation relations (\ref{quant1}) (or (\ref{quant2}))
and (\ref{quant3}).
Thus $I_{k_L,k_R}$ must be a $c$-number.
We can further show that $I_{k_L,k_R}$ satisfies the following relation:
%%%%%%%% 5.14
\beq
I_{k_L,k_R}I_{k'_L,k'_R} = I_{k_L+k'_L,k_R+k'_R}\ .
\eeq
%%%%%%%%%%%%%%
Hence $I_{k_L,k_R}$ would be of the form
%%%%%%%%%% 5.15
\beq
I_{k_L,k_R} = \exp\{ik_L\cdot a_L - ik_R\cdot a_R\}\ ,
\eeq
%%%%%%%%%%%%%%%%
for some constant vectors $a^I_L$ and $a^I_R$.
It seems that there is no way to determine the constant vectors
$a^I_L$ and $a^I_R$.
In the following, we will assume that
%%%%%%% 5.16
\beq
a^I_L = a^I_R  = 0\ .
\label{a=0}
\eeq
%%%%%%%%%%%%%%%%
We will, however, see that the above choice is consistent with
other requirements of string theory, in particular, with
modular invariance of partition functions.

A physical meaning of the identity (\ref{identity}) is obvious for
$[U,B] = 0$.
The identity (\ref{identity}) then reduces to
%%%%%%%%% 5.17
\beq
\exp\{i\pi k^I\widehat{w}^I\} \exp\{i\pi w^IU^{IJ}\widehat{p}'^J\}
= 1\ ,
\label{momquant0}
\eeq
%%%%%%%%%%%%%%%
where
%%%%%%%%% 5.18
\beqs
w^I &=& k^I_L - k^I_R\  \in\ \Lambda\ ,\nonumber\\
k^I &=& (1+B)^{IJ}k^J_L + (1-B)^{IJ}k^J_R\ \in\ 2\Lambda^{*}\ ,
        \nonumber\\
\widehat{p}'^I &=&
(\widehat{p}^a,-\frac{1}{\pi}(1-\widetilde{U}^T)^
  {\alpha\beta}\widehat{Q}^\beta)\ .
\eeqs
%%%%%%%%%%%%%%
The $\widehat{p}'^I$ is the canonical momentum conjugate to
$\widehat{x}^I$ in
the $g$-sector
\cite{Inoue-NT}.
Eq. (\ref{momquant0}) implies that
%%%%%%%% 5.20
\beqs
\widehat{w}^I &\in& \Lambda\ ,\nonumber\\
\widehat{p}'^I &\in& 2\Lambda^{*}\ .
\label{momquant1}
\eeqs
%%%%%%%%%%%%%%
These results are consistent with the identification
$x^I \sim x^I+\pi w^I$ for all $w^I \in \Lambda.$
For $[U,B]\ne 0,$ the first relation of eqs. (\ref{momquant1})
still holds while the second one does not.
In fact, putting $k^I=0$ and $w^I \in \Lambda_U$ in the identity
(\ref{identity}),
we have
%%%%%%%%% 5.21
\beq
\exp\Bigl\{i\pi w^I\widehat{p}^I_{_{/\! /}}
    - i\frac{\pi}{2}w^IC^{IJ}_gw^J\Bigr\}
= 1\ .
\eeq
%%%%%%%%%%%%%%%
This relation is equivalent to eq. (\ref{AB}) or (\ref{momshift}).

We should make a comment on the assumption (\ref{a=0}).
Using the identity (\ref{identity}), we can show
%%%%%%%
\beq
gV(k_L,k_R;z)g^{\dag} = V(k_L,k_R;e^{2\pi i}z)\ .
\eeq
%%%%%%%%%%%%
This is a desired relation because a $g$-invariant operator
$\sum^{N}_{m=1}g^mV(k_L,k_R;z)g^{-m}$ should be single-valued
with respect to $z$ although $V(k_L,k_R;z)$ itself may not be
single-valued in general. This fact supports the assumption
(\ref{a=0}).

%%%%%%%%%%%%%%%%%%%%%%%%%%%%%%%%%%%%%%%%%%%%%%%%%%%%%%%%%%%%%%%%%%%%%

\subsection{The Complete Set and the Inner Product of the
             Zero Mode Eigenstates}

In this subsection, we shall explicitly construct the complete
set and the inner product of the zero mode eigenstates.
The construction is not straightforward because the zero modes
obey the nontrivial commutation relations (\ref{quant2}) and
(\ref{quant3}).
We will see that the identity (\ref{identity}) derived in the
previous subsection plays a crucial role in the construction of the
complete set and the inner product of the zero mode eigenstates.

In the $g$-sector, the zero mode operators consist of
$\widehat{x}^I, \widehat{Q}^I, \widehat{p}^a$ and $\widehat{w}^I$
(or $\widehat{x}^I_L, \widehat{x}^I_R, \widehat{p}^a_L$ and
$\widehat{p}^a_R)$
with the constraint (\ref{fixedpoint}).
These operators obey the commutation relations (\ref{quant2}) and
(\ref{quant3}).
To construct the Hilbert space of the zero modes, we need to find
a maximal set of the zero mode operators which commute with
each other.
We will choose the following maximal set:
%%%%%%%%% 5/22
\beq
\{\ \widehat{p}^a\ \mbox{and}\ \widehat{w}^I\ ;\ a=1,\cdots,d,\
 I=1,\cdots,D\ \}\ .
\eeq
%%%%%%%%%%%%%
Let $|k^a_0, w^I_0>$ be an eigenstate of $\widehat{p}^a$ and
$\widehat{w}^I$.
Since the relevant operators are
$P^I_L(z), P^I_R(\overline{z})$ and $V(k_L,k_R;z)$ with
$(k_L,k_R) \in \Gamma^{D,D}$ and only the vertex operator can change
the eigenvalues of $\widehat{p}^a$ and $\widehat{w}^I$, other zero mode
eigenstates may be constructed by multiplying $|k^a_0, w^I_0>$ by
(the zero mode part of)
the vertex operator, i.e.
%%%%%%%%%%
\beq
\exp\{ik^I_L\widehat{x}^I_L + ik^I_R\widehat{x}^I_R\}
\ |k^a_0, w^I_0> \quad \mbox{for}\ (k^I_L,k^I_R) \in \Gamma^{D,D}\ ,
\nonumber\\
\eeq
%%%%%%%%%%%%%%%
or equivalently,
%%%%%%%%% 5.23
\beq
\exp\{ik^I\widehat{x}^I + iw^I\widehat{Q}^I\}
\ |k^a_0, w^I_0> \quad \mbox{for}\ k^I \in 2\Lambda^{*}\
\mbox{and}\ w^I \in \Lambda\ .
\label{state}
\eeq
%%%%%%%%%%%%%%%
Since the state (\ref{state}) is an eigenstate of $\widehat{p}^a$
and $\widehat{w}^I$ with the eigenvalues of $k^a + k^a_0$ and
$w^I + w^I_0$, respectively, we might conclude that
%%%%%%%%%% 5.25
\beq
(\widehat{p}^a,\widehat{w}^I) \in (2\Lambda^{*}|_{_{/\! /}}+k^a_0,
\Lambda+w^I_0)\ ,
\label{eigenv}
\eeq
%%%%%%%%%%%%
where
%%%%%%%% 5.26
\beq
2\Lambda^{*}|_{_{/\! /}} = \{\ k^a\ |\ k^I=(k^a,k^\alpha)
\in 2\Lambda^{*}\ \}\ .
\eeq
%%%%%%%%%%%%
We define the eigenstate $|k^a+k^a_0, w^I+w^I_0>$ by
%%%%%% 5.27
\beqs
|k^a+k^a_0,w^I+w^I_0>
&\equiv& \exp\{ik^a\widehat{x}^a+iw^I\widehat{Q}^I\}\ |k^a_0,w^I_0>
    \nonumber\\
&=& \zeta_{k,w}
    \exp\{ik^I\widehat{x}^I+iw^I\widehat{Q}^I\}\ |k^a_0,w^I_0>\ ,
\label{eigenstate}
\eeqs
%%%%%%%%%%%%%
where $k^I \in 2\Lambda^{*}, w^I \in \Lambda$ and
%%%%%%%% 5.28
\beq
\zeta_{k,w} = \exp\Bigl\{-i\frac{\pi}{2}k^\alpha
  \Bigl(\frac{1}{1-\widetilde{U}}\Bigr)^{\alpha\beta}
       (w^\beta+2w^\beta_0)\Bigr\}\ .
\eeq
%%%%%%%%%%%%%

So far, $k^a_0$ and $w^I_0$ are arbitrary.
To determine the allowed values of $k^a_0$ and $w^I_0$, let us
consider the identity (\ref{identity}).
For $w^I=k^I_L-k^I_R \in \Lambda_U,$ the identity (\ref{identity})
gives constraints on the eigenvalues of $\widehat{p}^a$ and
$\widehat{w}^I$.
It is easy to show that for $w^I \in \Lambda_U$ the identity
(\ref{identity}) reduces to
%%%%%%% 5.29
\beq
\exp\Bigl\{-i\frac{\pi}{2}w^IC^{IJ}_gw^J + i\pi w^a\widehat{p}^a
+ i\pi k^I\widehat{w}^I \Bigr\}\ =\ 1\ ,
\eeq
%%%%%%%%%%%%%
for all
$w^I\in \Lambda_U$ and $k^I \in 2\Lambda^{*}$.
This means that the eigenvalues of $\widehat{p}^a\
(\widehat{w}^I)$ must
belong to $2{\Lambda_U}^{*}+s^a_g\ (\Lambda)$, where we have used
the relation (\ref{sg}).
Taking account of the previous result (\ref{eigenv})
and using the fact that ${\Lambda_U}^{*} = \Lambda^{*}|_{_{/\!/}}$,
we conclude that without loss of generality we can set $k^a_0$ and
$w^I_0$ to be
%%%%%%%%%% 5.30
\beq
(k^a_0,w^I_0) = (s^a_g,0)\ .
\eeq
%%%%%%%%%%%%%

It is important to realize that all states $|k^a+s^a_g,w^I>$
with $k^a \in 2\Lambda^{*}|_{_{/\! /}}$ and $w^I \in \Lambda$ are
not independent.
To see this, let us consider the identity (\ref{identity}) again.
Multiplying $|k^a+s^a_g,w^I>$ by $I_{k'_L,k'_R}$, we find the
following relation:
%%%%%%%%% 5.31
\beq
|k^a+s^a_g+[U,B]^{aJ}w'^J, w^I+(1-U)^{IJ}w'^J >
= \rho(k,w;w')\ |k^a+s^a_g, w^I >\ ,
\label{momid}
\eeq
%%%%%%%%%%%%%
where $w'^I \equiv -(U^T)^{IJ}(k'^J_L-k'^J_R) \in \Lambda$ and
%%%%%%%%% 5.32
\beqs
\lefteqn{
 \rho(k,w;w')} \nonumber\\
& & =
   \exp\Bigl\{i\frac{\pi}{2}w'^IC^{IJ}_gw'^J + i\pi w'^a(k^a+s^a_g)
   + i\frac{\pi}{2}w'^I(U^T)^{IJ}B^{Ja}w^a
   - i\frac{\pi}{2}w'^aB^{aI}w^I \nonumber\\
& &\qquad +
   i\frac{\pi}{2}w'^I(B-U^TBU)^{I\alpha}\Bigl(\frac{1}{1-\widetilde{U}}
   \Bigr)^{\alpha\beta}w^\beta
   - i\frac{\pi}{2}w'^I(B-U^TBU)^{Ia}w'^a \Bigr\}\ .
\label{rho}
\eeqs
%%%%%%%%%%%%%%%%%
Thus we must identify the state in the left hand side of eq.
(\ref{momid}) with $|k^a+s^a_g,w^I>$ for all $w'^I \in \Lambda.$
(For $w'^I \in \Lambda_U,$ eq. (\ref{momid}) reduces to a trivial
identity.)
A geometrical meaning of the identification (\ref{momid}) is clear
for $[U,B]=0$:
Since a point $\{x^I\}$ is identified with $\{x^I+\pi w'^I\}$ for all
$w'^I \in \Lambda$ and
since $w^\alpha$ is related to $x^\alpha$ through the constraint
(\ref{fixedpoint}), $w^I$ should be identified with
$w^I+(1-U)^{IJ}w'^J$ for all $w'^I \in \Lambda.$
For $[U,B]\ne 0,$ a geometrical meaning of the identification
(\ref{momid}) is, however, less clear.

To obtain the complete set of the zero modes, let us first normalize
the state $|s^a_g,w^I=0>$ such that
$<s^a_g,w^I=0|s^a_g,w^I=0>\: = 1.$
Then, it follows that
%%%%%%%% 5.33
\beq
<k^a+s^a_g,w^I|k^a+s^a_g,w^I>\: = 1\ ,
\eeq
%%%%%%%%%%%%%
for all $k^a \in 2\Lambda^{*}|_{_{/\! /}}$ and $w^I \in \Lambda.$
Since the states
$|k^a+s^a_g+[U,B]^{aI}w'^I, w^I+(1-U)^{IJ}w'^J >$ must be identified
with $|k^a+s^a_g, w^I >$ for all $w'^I \in \Lambda,$ the complete set
of the zero modes is given by
%%%%%%%%% 5.34
\beq
1 = \sum_{k^a\in 2\Lambda^{*}|_{_{/\! /}}} \sum_{w^I \in \Lambda
/(1-U)\Lambda}
|k^a+s^a_g, w^I><k^a+s^a_g, w^I |\ ,
\label{completeset}
\eeq
%%%%%%%%%%%%%%
where $\Lambda/(1-U)\Lambda$ denotes the set of the independent lattice
points of $\Lambda$ with the identification
$w^I \sim w^I+(1-U)^{IJ}w'^J$ for all $w'^I \in \Lambda.$
To obtain the inner product of the zero mode eigenstates, we first
note that the inner product
$<k^a+s^a_g, w^I |\widetilde{k}^a+s^a_g, \widetilde{w} >$ can be nonzero
if and only if
%%%%%%% 5.35
\beqs
\widetilde{k}^a &=& k^a + [U,B]^{aI}w'^I\ ,\nonumber\\
\widetilde{w}^I &=& w^I + (1-U)^{IJ}w'^J\ ,
\eeqs
%%%%%%%%%%%%%%%
for some $w'^I \in \Lambda$ or precisely for some $w'^I \in
\Lambda/\Lambda_U.$
It turns out that
%%%%%%%% 5.36
\beq
<k^a+s^a_g, w^I | \widetilde{k}+s^a_g, \widetilde{w}^I >
= \sum_{w'^I \in \Lambda/\Lambda_U}
  \rho(k,w;w')\ \delta_{\widetilde{k}^a,k^a+[U,B]^{aI}w'^I}
  \delta_{\widetilde{w}^I,w^I+(1-U)^{IJ}w'^J}\ ,
\label{innerpro}
\eeq
%%%%%%%%%%%%%
where $\rho(k,w;w')$ is defined in eq. (\ref{rho}).
Since the summation of $w'^I$ is taken over $\Lambda/\Lambda_U,$
$\rho(k,w;w')$ should satisfy the following relation:
%%%%%%%% 5.37
\beq
\rho(k,w;w'+\Delta w) = \rho(k,w;w')\ ,
\qquad \mbox{for all}\ \Delta w^I \in \Lambda_U\ .
\label{rho1}
\eeq
%%%%%%%%%%%
To verify the relation (\ref{rho1}), we will use the relations
(\ref{Cg}) and (\ref{sg}).
The complete set (\ref{completeset}) and the inner product
(\ref{innerpro}) of the zero mode eigenstates will be used to compute
one loop partition functions
in the next section.

Finally we discuss the degeneracy of the ground state in the
$g$-sector, which plays an important role in the determination
of the massless spectrum.
To this end, it may be convenient to rewrite $|k^a+s^a_g, w^I>$
in terms of $(k^a_L,k^a_R)$ and the fixed \lq\lq point"
$x^\alpha$ as follows:
%%%%%% 5,39
\beq
|\:k^a_L+\frac{1}{2}s^a_g, k^a_R+\frac{1}{2}s^a_g; x^\alpha >
\equiv |\:k^a+s^a_g,w^I >\ ,
\eeq
%%%%%%%%%%%%%%
where $k^a_L+s^a_g/2$ ($k^a_R+s^a_g/2$) corresponds to the
eigenvalue of ${\widehat p}^a_L$ (${\widehat p}^a_R$) and
%%%%%%%%% 5.40
\beqs
k^a_L &=& \frac{1}{2}(k^a-B^{aI}w^I) + \frac{1}{2}w^a\ ,\nonumber\\
k^a_R &=& \frac{1}{2}(k^a-B^{aI}w^I) - \frac{1}{2}w^a\ ,\nonumber\\
x^\alpha
      &=& \pi \Bigl(\frac{1}{1-\widetilde{U}}\Bigr)^{\alpha\beta}w^\beta\ .
\label{kaLkaRx}
\eeqs
%%%%%%%%%%%%%
In this notation, the identification (\ref{momid}) can simply be
written as
%%%%%%%%% 5.41
\beq
|\:k^a_L+\frac{1}{2}s^a_g,k^a_R+\frac{1}{2}s^a_g;x^\alpha + \pi w'^\alpha >
    \ \sim |\:k^a_L+\frac{1}{2}s^a_g,k^a_R+\frac{1}{2}s^a_g;x^\alpha >
    \quad \mbox{for all}\ w'^\alpha \in \Lambda|_{\perp}\ ,
\eeq
%%%%%%%%%%%%%%
where
%%%%%%%%
\beq
\Lambda|_{\perp} \equiv \{w'^\alpha\ |\ w'^I=(w'^a,w'^\alpha)
\in \Lambda\ \}\ .
\eeq
%%%%%%%%%%%%%
The allowed values of $(k^a_L,k^a_R)$ are given by
%%%%%%%%% 5.42
\beq
(k^a_L,k^a_R) \in \Gamma^{D,D}|_{_{/\! /}} \equiv
\{\ (k^a_L,k^a_R)\ |\ (k^I_L,k^I_R) \in \Gamma^{D,D}\ \}\ .
\eeq
%%%%%%%%%%%%%
Then, the eigenvalues of the fixed \lq\lq point" $x^\alpha$
are given by
%%%%%%%%% 5.43
\beq
x^\alpha = \pi \Bigl(\frac{1}{1-\widetilde{U}}\Bigr)^
             {\alpha\beta}w^\beta\ ,
\eeq
%%%%%%%%%%%%%
where
%%%%%%%%% 5.44
\beq
w^\alpha \in\frac{\Lambda|_{\perp}}{(1-U)\Lambda|_{\perp}}\qquad
\mbox{with}\ (k^a_L-k^a_R, w^\alpha) \in \Lambda\ .
\eeq
%%%%%%%%%%%%%%%
The number of the fixed \lq\lq points" is equal to the number of the
different eigenvalues of $x^\alpha$ with a fixed $(k^a_L,k^a_R),$
i.e.
\cite{asymorb1}
%%%%%%%%% 5.45
\beq
\left| \frac{\Lambda^{inv}_{\perp}}{(1-U)\Lambda|_{\perp}}
\right| \ ,
\label{nofixed}
\eeq
%%%%%%%%%%%%%%
where
%%%%%%
\beq
\Lambda^{inv}_{\perp} = \left\{\ w^\alpha\ |\ w^I=(w^a=0,w^\alpha)
   \in \Lambda\ \right\}\ .
\eeq
%%%%%%%%%%%%%%
The formula (\ref{nofixed}) can also be represented by
%%%%%%%% 5.46
\beq
\frac{\det(1-\widetilde{U})}{V_{\Gamma_U}}\ ,
\label{nofixed'}
\eeq
%%%%%%%%%%%%%%
where $V_{\Gamma_U}$ denotes the volume of the unit cell of the lattice
$\Gamma_U$ and
%%%%%%%% 5.47
\beq
\Gamma_U = \{\ (k^a_L,k^a_R)\ |\ (k^I_L,k^I_R) \in \Gamma^{D,D}
            \ \mbox{with}\ (k^\alpha_L,k^\alpha_R)=0\ \}\ .
\eeq
%%%%%%%%%%%%%%%%%
It should be emphasized that the degeneracy of the ground state
is {\it not\/}, in general, identical to the number of the fixed
\lq\lq points" (\ref{nofixed}) or (\ref{nofixed'}) due to the
momentum degeneracy.
If $s^a_g \notin 2\Lambda^{*}|_{_{/\!/}}$ or equivalently if
$(s^a_g/2,s^a_g/2) \notin \Gamma^{D,D}|_{_{/\!/}}$,
$(k^a_L+s^a_g/2,k^a_R+s^a_g/2)$
cannot be equal to $(0,0)$ and hence the momentum
degeneracy may occur.
Let $n_{_M}$ be the number of the different momentum eigenvalues
with the minimum value
of $(k^a_L+s^a_g/2)^2 + (k^a_R+s^a_g/2)^2.$
Then, the degeneracy of the ground state in the $g$-sector
is given by
%%%%%%%
\beq
n_{_M} \times \frac{\det(1-\widetilde{U})}{V_{\Gamma_U}}\ .
\eeq
%%%%%%%%%%%%

\vspace{1cm}

%%%%%%%%%%%%%%%%%%%%%%%%%%%%%%%%%%%%%%%%%%%%%%%%%%%%%%%%%%%%%%%%%%%%%%%%%
%%%%%%%%%%%%%%%%%%%%%%%%%%%%%%%%%%%%%%%%%%%%%%%%%%%%%%%%%%%%%%%%%%%%%%%%%

\sect{One Loop Modular Invariance}

In this section, we shall prove one loop modular invariance of
partition functions.
Let $Z(g,h;\tau)$ be the partition function of the $g$-sector twisted
by $h$ which is defined, in the operator formalism, by
%%%%%%%%% 7.1
\beq
Z(g,h;\tau) =
{\rm Tr}\Bigl[\:h\:e^{i2\pi \tau(L_0-D/24)-i2\pi \overline{\tau}
    (\overline{L}_0-D/24)}\:\Bigr]_{g-sector}\ ,
\label{partitiongh}
\eeq
%%%%%%%%%%%%%
where $L_0$ $(\overline{L}_0)$ is the Virasoro zero mode operator of the
left- (right-)mover.
The trace in eq. (\ref{partitiongh}) is taken over the Hilbert space
of the $g$-sector.
Then, the one loop partition function will be of the form,
%%%%%%%%%% 7.2
\beq
Z(\tau) = \frac{1}{|G|}\sum_{{g,h \in G}\atop{gh=hg}}Z(g,h;\tau)\ ,
\label{partition}
\eeq
%%%%%%%%%%%%%%%
where $|G|$ is the order of the group $G$.
In the above summation, only the elements $g$ and $h$ which commute
with each other contribute to the partition function.
This will be explained as follows:
Let ${\cal H}_g$ be the Hilbert space of the $g$-sector.
The total Hilbert space ${\cal H}_{total}$ consists of the direct
sum of ${\cal H}_g$ for all $g \in G,$
%%%%%%%%%%%% 7.3
\beq
{\cal H}_{total} = \bigoplus_{g\in G} {\cal H}_g\ .
\eeq
%%%%%%%%%%%%%%
The physical Hilbert space is not the total Hilbert space but the
$G$-invariant subspace of ${\cal H}_{total}.$
Thus, the partition function will be given by
%%%%%%%%% 7.4
\beq
Z(\tau) = \sum_{g\in G} Z(\tau)_{g-sector}\ ,
\eeq
%%%%%%%%%%%
where
%%%%%%%% 7.5
\beq
Z(\tau)_{g-sector} = {{\rm Tr}}^{(phys)}
    \Bigl[\:e^{i2\pi \tau(L_0-D/24)
    -i2\pi \overline{\tau}(\overline{L}_0-D/24)}\:\Bigr]_{g-sector}\ .
\label{partitiong}
\eeq
%%%%%%%%%%%%
Here, the trace should be taken over the physical Hilbert space of the
$g$-sector, which will be given by
%%%%%%%% 7.6
\beq
{\cal H}^{(phys)}_g = {\cal P} {\cal H}_g\ ,
\eeq
%%%%%%%%%%%%%
where ${\cal P}$ is the projection operator defined by
%%%%%%%%% 7.7
\beq
{\cal P} = \frac{1}{|G|} \sum_{h\in G} h\ .
\eeq
%%%%%%%%%%%%%
By use of the projection operator ${\cal P},$ the trace formula
(\ref{partitiong}) can be rewritten as
%%%%%%%% 7.8
\beq
Z(\tau)_{g-sector} = {\rm Tr}\Bigl[\:{\cal P}\:e^{i2\pi \tau(L_0-D/24)
    -i2\pi \overline{\tau}(\overline{L}_0-D/24)}\:\Bigr]_{g-sector}\ ,
\label{partitiong'}
\eeq
%%%%%%%%%%%%
where the trace is taken over the Hilbert space ${\cal H}_g.$
The twist operator $h$ maps ${\cal H}_g$ onto
${\cal H}_{hgh^{-1}}$.
%%%%%%%%% 7.9
\beq
h : \quad {\cal H}_g\ \rightarrow\ {\cal H}_{hgh^{-1}}\ .
\eeq
%%%%%%%%%%%
Therefore, in the trace formula (\ref{partitiong'}),
%%%%%%% 7.10
\beq
{\rm Tr}\left[\:h\:
    e^{i2\pi \tau(L_0-D/24)
       -i2\pi \overline{\tau}(\overline{L}_0-D/24)}\:\right]
_{g-sector}\ ,
\eeq
%%%%%%%%%%%%%
will vanish identically unless $h$ commutes with $g.$

The modular group is generated by two transformations
$T:\tau \to \tau+1$ and $S:\tau \to -1/\tau$.
Thus, one loop modular invariance of the partition function
(\ref{partition}) is satisfied provided
%%%%%%%% 7.11
\beqs
Z(g,h;\tau+1) &=& Z(g,gh;\tau)\ ,
\label{modultr1}    \\
Z(g,h;-1/\tau) &=& Z(h^{-1},g;\tau)\ .
\label{modultr2}
\eeqs
%%%%%%%%%%%%
Let us calculate $Z(g,h;\tau)$ defined in eq. (\ref{partitiongh}).
The $g$ and $h$ are assumed to commute with each other and to be
associated with the orthogonal matrices $U^{IJ}$ and $V^{IJ}$,
respectively.
The condition $gh = hg$ is equivalent to the following two conditions:
%%%%%%% 7.14
\beq
UV\ =\ VU\ ,
\label{UVVU}
\eeq
%%%%%%%%%%%
and
%%%%%% 7.15
\beq
\frac{1}{2}w^I\left(UC_gU^T + UVC_hV^TU^T
  - VC_hV^T - VUC_gU^TV^T\right)^{IJ}
  w^J = 0\quad \mbox{mod}\ 2\  ,
\label{condCU}
\eeq
%%%%%%%%%%%%
for all $w^I \in \Lambda$.
The second condition (\ref{condCU}) comes from the consistency with
eq. (\ref{hV}).
The two conditions (\ref{UVVU}) and (\ref{condCU}) also assure
the consistency with the identity (\ref{identity}).
Since $U^{IJ}$ commutes with $V^{IJ}$, we can take a coordinate
system for $X^I$ in such a way that $U^{IJ}$ and $V^{IJ}$ are
diagonal with respect to their invariant subspaces as follows:
%%%%%%%%% 7.16
\beqs
U^{IJ} &=&
  \left(
  \begin{array}{cccc}
    \delta^{i_1j_1} & 0 & 0 & 0 \\
    0 & \delta^{i_2j_2} & 0 & 0 \\
    0 & 0 & u^{i_3j_3}_3 & 0   \\
    0 & 0 & 0 & u^{i_4j_4}_4
  \end{array}
  \right)^{IJ}\ ,\nonumber\\
V^{IJ} &=&
  \left(
  \begin{array}{cccc}
    \delta^{i_1j_1} & 0 & 0 & 0 \\
    0 & v^{i_2j_2}_2 & 0 & 0 \\
    0 & 0 & \delta^{i_3j_3} & 0   \\
    0 & 0 & 0 & v^{i_4j_4}_4
  \end{array}
  \right)^{IJ}\ ,\nonumber\\
X^I &=& ( X^{i_1},X^{i_2},X^{i_3},X^{i_4} )\ ,
\label{UVsystem}
\eeqs
%%%%%%%%%%%%%%%
where $i_k,j_k = 1,2,\cdots,d_k (k=1,2,3,4)$ and
$I,J=1,2,\cdots,D\ (D=d_1+d_2+d_3+d_4).$
The correspondence with the coordinate system (\ref{cosystem})
will be apparent.
Let $N$ and $N'$ be the smallest positive integers such that
$g^N=1$ and $h^{N'}=1.$
Then, it follows that
%%%%%%% 7.18
\beqs
U^N &=& {\bf 1}\ ,\nonumber\\
V^{N'} &=& {\bf 1}\ ,
\eeqs
%%%%%%%%%%%
and
%%%%%%%%% 7.19
\beqs
\sum^{N}_{m=1} \frac{1}{2}w^I(U^mC_gU^{-m})^{IJ}w^J &=& 0
\quad \mbox{mod}\ 2\ ,\nonumber\\
\sum^{N'}_{m=1} \frac{1}{2}w^I(V^mC_hV^{-m})^{IJ}w^J &=& 0
\quad \mbox{mod}\ 2\ ,\qquad \mbox{for all}\ w^I \in \Lambda\ .
\label{gN=1}
\eeqs
%%%%%%%%%%%%%
The conditions (\ref{gN=1}) come from the consistency with eq.
(\ref{hV}).
Since $U^{IJ}$ commutes with $V^{IJ},$
the orthogonal matrices $U^{IJ}$ and $V^{IJ}$ can be
diagonalized simultaneously by a single unitary matrix $M^{IJ}$:
%%%%%%%% 7.22
\beqs
M^{\dag}UM &=& U_{diag}\ ,\nonumber\\
M^{\dag}VM &=& V_{diag}\ ,
\eeqs
%%%%%%%%%%%%%%
where
%%%%%%%7.23
\beqs
U^{IJ}_{diag} &=& \omega^{t_I}\ \delta^{IJ}\
      \qquad(\omega=e^{i2\pi/N})\ ,\nonumber\\
V^{IJ}_{diag} &=& \omega'^{t'_I}\ \delta^{IJ}\
      \qquad(\omega'=e^{i2\pi/N'})\ .
\eeqs
%%%%%%%%%%%%%%
Here, we choose $t_I$ and $t'_I$ such that
$0 \le t_I \le N-1$ and $0 \le t'_I \le N'-1.$
Since we have taken the coordinate system (\ref{UVsystem}), we may
have
%%%%% 7.25
\beq
t_{i_1} = t_{i_2} = t'_{i_1} = t'_{i_3} = 0\ .
\eeq
%%%%%%%%%%%%
The Virasoro zero mode operators $L_0$ and $\overline{L}_0$
in the $g$-sector are given by
%%%%%%%
\beqs
L_0 &=& \frac{1}{2}\sum^{d}_{a=1}(\widehat{p}^a_L)^2
        + \sum^{D}_{I=1}\sum_{n_I\in{\bf Z}-t_I/N>0}
         {\gamma^I_{n_I}}^{\dag}{\gamma^I_{n_I}} + \varepsilon_g\ ,
         \nonumber\\
\overline{L}_0 &=& \frac{1}{2}\sum^{d}_{a=1}(\widehat{p}^a_R)^2
        + \sum^{D}_{I=1}\sum_{n_I\in{\bf Z}+t_I/N>0}
         {\overline{\gamma}^I_{n_I}}^{\dag}
         {\overline{\gamma}^I_{n_I}} + \varepsilon_g\ .
\eeqs
%%%%%%%%%%%%%%
The $\varepsilon_g$ is the conformal weight of the ground state
in the $g$-sector
\cite{orbifold},
%%%%%%%
\beq
\varepsilon_g = \frac{1}{4}\sum^{D}_{I=1}\frac{t_I}{N}
                \left(1-\frac{t_I}{N}\right)\ .
\eeq
%%%%%%%%%%%%%%%
It is convenient to calculate the zero mode part and the oscillator
part of $Z(g,h;\tau)$ separately.
%%%%%%%% 7.26
\beq
Z(g,h;\tau) = Z(g,h;\tau)_{zero}\times Z(g,h;\tau)_{osc}\ ,
\eeq
%%%%%%%%%%%
where
%%%%%%% 7.27
\beqs
Z(g,h;\tau)_{zero}
  &=& {\rm Tr}\Bigl[\: h_{zero}\: e^{i\pi\tau(\widehat{p}^a_L)^2
        - i\pi\overline{\tau}({\widehat p}^a_R)^2}\:\Bigr]
         ^{zero}_{g-sector}\ ,\nonumber\\
Z(g,h;\tau)_{osc}
  &=& e^{-\pi{\rm Im}\tau(4\varepsilon_g-D/6)}
       {\rm Tr}\Bigl[\: h_{osc}\:
       e^{i2\pi\tau\sum\gamma^{I\dag}_{n_I}\gamma^I_{n_I}
        - i2\pi\overline{\tau}\sum \overline{\gamma}^{I\dag}_{n_I}
       \overline{\gamma}^I_{n_I}}\: \Bigr]^{osc}_{g-sector}\ .
\eeqs
%%%%%%%%%%%%%%

Let us first calculate $Z(g,h;\tau)_{osc}$.
The twist operator $h$ acts on the oscillator part of
$(X^I_L(z),X^I_R(\overline{z}))$ as
%%%%%% 7.29
\beq
h :\quad (X^I_L(z),X^I_R(\overline{z}))_{osc}
  \ \rightarrow\ (V^{IJ}X^J_L(z),V^{IJ}X^J_R(\overline{z}))_
  {osc}\ .
\eeq
%%%%%%%%%%%%
This means that the action of $h$ on $\gamma^I_{n_I}$ and
$\overline{\gamma}^I_{n_I}$ is given by
%%%%%%%%% 7.30
\beqs
h :\quad \gamma^I_{n_I} &\rightarrow& \omega'^{t'_I}\
         \gamma^I_{n_I}\ , \nonumber\\
         \overline{\gamma}^I_{n_I} &\rightarrow& \omega'^{t'_I}\
         \overline{\gamma}^I_{n_I}\ .
\eeqs
%%%%%%%%%%%%%%%
The $h$ acts trivially on the Fock vacuum of the oscillators.
The calculation of $Z(g,h;\tau)_{osc}$ is straightforward
and the result is
%%%%%%%%%% 7.31
\beq
Z(g,h;\tau)_{osc}
= \prod^{D}_{I=1}\ \prod_{n_I\in {\bf Z}-t_I/N >0}
   (1-\omega'^{-t'_I}e^{i2\pi\tau n_I})^{-1}
   \prod_{n_I\in{\bf Z}+t_{I}/N>0}
   (1-\omega'^{-t'_I}e^{-i2\pi\overline{\tau} n_I})^{-1}\ .
\eeq
%%%%%%%%%%%%%
In terms of the theta functions, $Z(g,h;\tau)_{osc}$
can be written as
%%%%% 7.32
\beq
Z(g,h;\tau)_{osc}
 = \frac{\det(1-v_2)}{|\eta(\tau)|^{2d_1}}
   \prod_{k=2,3,4}
\prod^{d_k}_{{i_k}=1} \left|\frac{\eta(\tau)}
 {e^{i\pi\tau(t_{i_k}/N)^2}
  \vartheta_1(t_{i_k}\tau/N+t'_{i_k}/N'|\tau)}\right|\ ,
\label{partosc}
\eeq
%%%%%%%%%%%
where we have used the fact that the set of the eigenvalues
$\{\omega^{t_I}, \omega'^{t'_I}; I=1,\cdots,D\}$
is identical to
$\{\omega^{-t_I}, \omega'^{-t'_I}; I=1,\cdots,D\}$.
The functions $\eta(\tau)$  and $\vartheta_a(\nu|\tau)$ are
the Dedekind $\eta$-function and the Jacobi theta function,
which are defined in appendix B.

Let us next calculate the zero mode part of $Z(g,h;\tau).$
Using the complete set (\ref{completeset}), $Z(g,h;\tau)_{zero}$
can be given by
%%%%%%% 7.33
\beqs
Z(g,h;\tau)_{zero}
  &=&  \sum_{k^a \in 2\Lambda^{*}|_{_{/\! /}}}
        \sum_{w^I \in \Lambda/(1-U)\Lambda}
        e^{i\pi\tau(k^a_L+s^a_g/2)^2
        - i\pi\overline{\tau}(k^a_R+s^a_g/2)^2}
        \nonumber\\
  & &\quad \times
        < k^a + s^a_g, w^I |\: h_{zero}\: |k^a + s^a_g, w^I >\ ,
\label{partitionzero}
\eeqs
%%%%%%%%%%%%%
where $k^a_L$ and $k^a_R$ are defined in eq. (\ref{kaLkaRx}).
To compute the inner product in eq. (\ref{partitionzero}),
we need to know the action of $h$ on $|s^a_g, w^I=0>,$ which
will be given by
%%%%%%%% 7.36
\beq
h :\quad |s^a_g, w^I=0>\ \ \rightarrow\
   \ |(\widetilde{V}^T)^{ab}s^b_g, w^I=0>\ .
\label{hground1}
\eeq
%%%%%%%%%%%%%
One might expect that a nontrivial phase would appear in the right
hand side of eq. (\ref{hground1}).
Modular invariance, however, requires no such phase, as we will see
later.
We would have problems if
$(\widetilde{V}^T)^{ab}s^b_g - s^a_g \notin 2\Lambda^{*}|_{_{/\! /}},$
because this means that the state $|(\widetilde{V}^T)^{ab}s^b_g,w^I=0>$
would not belong to ${\cal H}_g.$
However, this is not the case.
For $w^I \in \Lambda_U$, the condition (\ref{condCU}) reduces to
%%%%%%% 7.38
\beq
w^a\left(s^a_g - \widetilde{V}^{ab}s^b_g \right) = 0\quad \mbox{mod}\ 2\ .
\eeq
%%%%%%%%%%%
This guarantees that $h|s^a_g,w^I=0>$ belongs to ${\cal H}_g.$
Using eqs. (\ref{anomaltr2}), (\ref{eigenstate}), (\ref{innerpro}),
(\ref{partitionzero}) and (\ref{hground1}), we can now compute
$Z(g,h;\tau)_{zero}.$
The result is
%%%%%%%% 7.39
\beq
Z(g,h;\tau)_{zero}\ =\
   \sum_{k^a \in 2\Lambda^{*}|_{_{/\! /}}}
   \sum_{w^I \in \Lambda/(1-U)\Lambda}
   \sum_{w'^I \in \Lambda/\Lambda_U}
   {\cal A}(g,h;\tau;k^a,w^I,w'^I)\ ,
\label{partzero1}
\eeq
%%%%%%%%%%%%
where
%%%%%% 7.40
\beqs
\lefteqn{{\cal A}(g,h;\tau;k^a,w^I,w'^I)}\nonumber\\
 & & = \delta_{(1-\widetilde{V})^{ab}(k^b+s^b_g-B^{bI}w^I),0}
      \delta_{(1-V)^{IJ}w^J,(1-U)^{IJ}w'^J} \nonumber\\
 & & \quad \times
     \exp\Bigl\{i\pi\tau(k^a_L+\frac{1}{2}s^a_g)^2
              - i\pi\overline{\tau}(k^a_R+\frac{1}{2}s^a_g)^2
       -i\frac{\pi}{2}w^IC^{IJ}_hw^J - i\frac{\pi}{2}w'^IC^{IJ}_gw'^J
     \nonumber\\
 & &\qquad
     + i\frac{\pi}{2}w'^I(U^TBV)^{IJ}w^J
         + i\frac{\pi}{2}w'^IB^{IJ}w^J
         + i\pi w'^a (k^a + s^a_g - B^{aI}w^I) \Bigr\}\ .
\label{partzero1'}
\eeqs
%%%%%%%%%%%%%%
Here, we have appropriately redefined the variables.
We note that the function ${\cal A}$ satisfies the following equations:
%%%%%%% 7.41
\beqs
& &{\cal A}(g,h;\tau;k^a+B^{aI}(1-U)^{IJ}\Delta w^J,
            w^I+(1-U)^{IJ}\Delta w^J,
            w'^I+(1-V)^{IJ}\Delta w^J ) \nonumber\\
& & \qquad=\ {\cal A}(g,h;\tau;k^a,w^I,w'^I)\ ,\qquad \mbox{for all}\
  \Delta w^I \in \Lambda\ ,\nonumber\\
& & {\cal A}(g,h;\tau;k^a,w^I,w'^I+\Delta w'^I) \nonumber\\
& & \qquad=\ {\cal A}(g,h;\tau;k^a,w^I,w'^I)\ ,\qquad \mbox{for all}\
  \Delta w'^I \in \Lambda_U\ .
\eeqs
%%%%%%%%%%%%%%
These symmetries guarantee the well-definedness of the summations of
$w^I$ and $w'^I$ in eq. (\ref{partzero1}).

We now discuss the modular transformation property of $Z(g,h;\tau)$.
Let us first consider the relation (\ref{modultr1}).
Although it is not difficult to prove the relation (\ref{modultr1})
from the
expressions (\ref{partosc}), (\ref{partzero1}) and (\ref{partzero1'}),
we will
give a more direct proof of eq. (\ref{modultr1}) here.
The expression (\ref{partitiongh}) and the relation (\ref{modultr1})
suggest that the twist operator $g$ in the $g$-sector is
given by
%%%%%%% 7.44
\beq
g = \exp \left\{i2\pi (L_0-\overline{L}_0)\right\}\ .
\label{g}
\eeq
%%%%%%%%%%%%%
Indeed, we can verify that $g$ defined in eq. (\ref{g}) satisfies
all desired relations, in particular,
%%%%%%%% 7.45
\beq
g V(k_L,k_R;z) g^{\dag}
  = \exp \left\{-i\frac{\pi}{2}(k_L-k_R)^I(UC_gU^T)^{IJ}
    (k_L-k_R)^J\right\}\ V(U^Tk_L,U^Tk_R;z)\ .
\label{gVg}
\eeq
%%%%%%%%%%%%%%
To show eq. (\ref{gVg}), we will use the identity (\ref{identity}).
Thus, the above observation (\ref{g}) proves the relation (\ref{modultr1}).

It should be emphasized that the left-right level matching condition
\cite{asymorb1,Vafa}
%%%%%%%%%
\beq
Z(g,h;\tau+N) = Z(g,h;\tau)\ ,
\label{levelmatch'}
\eeq
%%%%%%%%%%%%%%
or equivalently,
%%%%%%%% 7.46
\beq
N\left( (k^a_L+\frac{1}{2}s^a_g)^2
         - (k^a_R+\frac{1}{2}s^a_g)^2 \right) = 0
   \quad  \mbox{mod}\ 2\ \quad \mbox{for all}\ (k^a_L,k^a_R)
   \in \Gamma^{D,D}|_{_{/\!/}}\ ,
\label{levelmatch}
\eeq
%%%%%%%%%%%%
would be destroyed in general if we put $s^a_g$ to be equal to
zero by hand.
The existence of the shift vector $s^a_g$ guarantees the condition
(\ref{levelmatch}) in general, as shown in ref.
\cite{Sakamoto-TH}.

Let us next consider the relation (\ref{modultr2}).
It is easy to show that
%%%%%%%%% 7.47
\beq
Z(g,h;-1/\tau)_{osc}
  = |\tau|^{-d_1}\frac{\det(1-v_2)}{\det(1-u_3)}
    Z(h^{-1},g;\tau)_{osc}\ .
\eeq
%%%%%%%%%%%%%%
This relation implies that the zero mode part of $Z(g,h;\tau)$ should
satisfy
%%%%%%% 7.48
\beq
Z(g,h;-1/\tau)_{zero}
    = |\tau|^{d_1}\frac{\det(1-u_3)}{\det(1-v_2)}
    Z(h^{-1},g;\tau)_{zero}\ .
\label{modulzero2}
\eeq
%%%%%%%%%%%%%%
The expressions (\ref{partzero1}) and (\ref{partzero1'})
may not be convenient to prove the
relation (\ref{modulzero2}).
The following expression of $Z(g,h;\tau)_{zero}$ will be
suitable to prove the relation (\ref{modulzero2}):
%%%%%%%%% 7.49
\beq
   Z(g,h;\tau)_{zero} =
    \frac{1}{\pi^{d_1}\det(1-v_2)}\left(\frac{1}{2\tau_2}\right)^{d_1/2}
    \int_{\pi \Lambda} dx^I
    \sum_{w^I\in \Lambda}
    \sum_{w'^I \in \Lambda}
    {\widetilde{\cal A}}(g,h;\tau;x^I,w^I,w'^I)\ ,
\label{pathint}
\eeq
%%%%%%%%%%%%%%%%%%
where $\tau = \tau_1+i\tau_2$ and
%%%%%%%% 7.50
\beqs
\lefteqn{{\widetilde{\cal A}}(g,h;\tau;x^I,w^I,w'^I)} \nonumber\\
& &
  =  \delta\Bigl(x^{i_2}
        - \pi \Bigl(\frac{1}{1-v_2}\Bigr)^{i_2j_2}w'^{j_2}\Bigr)
    \delta\Bigl(x^{\alpha}
        - \pi \Bigl(\frac{1}{1-\widetilde{U}}\Bigr)
      ^{\alpha\beta}w^{\beta}\Bigr)
    \delta_{(1-V)^{IJ}w^J,(1-U)^{IJ}w'^J}   \nonumber\\
& &\quad \times
    \exp\Bigl\{ -\frac{\pi}{2}
     \Bigl[\
       \frac{|\tau|^2}{\tau_2}(w^{i_1})^2 + 2\frac{\tau_1}{\tau_2}
       w^{i_1}w'^{i_1} + \frac{1}{\tau_2}(w'^{i_1})^2\
     \Bigr]  \nonumber\\
& & \quad\
       -i\frac{\pi}{2}w^IC^{IJ}_hw^J
       - i\frac{\pi}{2}w'^IC^{IJ}_gw'^J
       +i\frac{\pi}{2}w'^I(U^TBV)^{IJ}w^J
       +i\frac{\pi}{2}w'^IB^{IJ}w^J \Bigr\}.
\label{pathint'}
\eeqs
%%%%%%%%%%%%%%%
The derivation of eqs. (\ref{pathint}) and (\ref{pathint'})
will be given in appendix C.
Then, it is not difficult to show from eqs. (\ref{pathint})
and (\ref{pathint'}) that
%%%%%%%%%
\beqs
Z(g,h;-1/\tau)_{zero}
 &=&  \frac{1}{\pi^{d_1}\det(1-v_2)}\left(\frac{|\tau|^2}
      {2\tau_2}\right)^{d_1/2}
     \int_{\pi \Lambda} dx^I
     \sum_{w^I\in \Lambda}
     \sum_{w'^I \in \Lambda} \nonumber\\
 & &\qquad \times\
     {\widetilde{\cal A}}(h^{-1},g;\tau;x^I,-(V^T)^{IJ}w'^J,w^I)\ ,
\eeqs
%%%%%%%%%%%%%%%%%%
where $C^{IJ}_{h^{-1}}$ is related to $C^{IJ}_h$ as
%%%%%%
\beq
\frac{1}{2}w^IC^{IJ}_{h^{-1}}w^J =
\frac{1}{2}w^I(VC_hV^T)^{IJ}w^J\quad \mbox{mod 2}\ ,
\eeq
%%%%%%%%%%%%%%%%%
for all $w^I \in \Lambda.$
Replacing $-(V^T)^{IJ}w'^J$ and $w^I$ with $w^I$ and $w'^I$,
respectively,
we obtain the relation (\ref{modulzero2}).

\vspace{1cm}

%%%%%%%%%%%%%%%%%%%%%%%%%%%%%%%%%%%%%%%%%%%%%%%%%%%%%%%%%%%%%%%%%%%
%%%%%%%%%%%%%%%%%%%%%%%%%%%%%%%%%%%%%%%%%%%%%%%%%%%%%%%%%%%%%%%%%%%

\sect{An Example}

In this section, we shall investigate a $Z_2$ orbifold
model with a topologically nontrivial twist $g$ in detail,
which will give a good illustration of our formulation.

Let us introduce the root lattice $\Lambda_R$ and the weight lattice $
\Lambda_W$ of $SU(3)$ as
%%%%%%%
\beqs
\Lambda_R&=&\{\ q^I=\sum_{i=1}^2 n^i\alpha^I_i,\ n^i \in {\bf Z}\ \}\ ,
  \nonumber\\
\Lambda_W&=&\{\ q^I=\sum_{i=1}^2 m_i\mu^{iI},\ m_i \in {\bf Z}\ \}\ ,
\eeqs
%%%%%%%%%%%%%
where $\alpha_i$ and $\mu^i$ $(i=1,2)$ are a simple root and a fundamental
weight satisfying $\alpha_i\cdot\mu^j={\delta_i}^j$.
We will take $\alpha_i$
and $\mu^i$ to be
%%%%%%%%%
\beqs
\alpha^I_1 = ({1\over\sqrt 2},\sqrt{3\over2})\ ,& &
    \alpha^I_2 =  ({1\over\sqrt 2},-\sqrt{3\over2})\ ,\nonumber\\
\mu^{1I} = ({1\over\sqrt 2},\sqrt{1\over6})\ ,& &
    \mu^{2I} = ({1\over\sqrt 2},-\sqrt{1\over6})\ .
\eeqs
%%%%%%%%%%%%%
Note that $\Lambda_W$ is the dual lattice of $\Lambda_R.$
The center of mass momentum $p^I $and the winding
number $w^I$ are assumed to lie on the following lattices:
%%%%%%%%
\beqs
p^I &\in& 2\Lambda_W\ ,\nonumber\\
w^I &\in& \Lambda_R\ .
\eeqs
%%%%%%%%%%%%%%
The antisymmetric background field $B^{IJ}$ is chosen as
%%%%%%%%
\beq
B^{IJ}=\pmatrix{0 & -{1\over\sqrt{3}} \cr
                  {1\over\sqrt{3}} & 0 \cr}^{IJ}\ .
\eeq
%%%%%%%%%%%%%%%
Then, it turns out that
the left- and right-moving momentum $(p^I_L,p^I_R)$
lies on the following $(2+2)$-dimensional Lorentzian
even self-dual lattice $\Gamma^{2,2}$:
%%%%%%%%
\beq
\Gamma^{2,2}=\{\ (p^I_L,p^I_R)\ |\
               p^I_L,p^I_R\in \Lambda_W,\ p^I_L-p^I_R\in
               \Lambda_R\ \}\ .
\eeq
%%%%%%%%%%%%%%%
This is an example of the lattice (\ref{ENlattice}).
We consider the following $Z_2$ transformation:
%%%%%%%
\beq
g:\quad X^I\ \rightarrow\ U^{IJ}X^J\ ,
\label{Z2}
\eeq
%%%%%%%%%%%%%%%
where
%%%%%%%
\beq
U^{IJ} = \pmatrix{1 & 0 \cr
                             0  & -1 \cr }^{IJ}\ .
\eeq
%%%%%%%%%%%%%%%
This is an automorphism of $\Lambda_R$ and also $\Gamma^{2,2}$,
as it should be.
According to our prescription, we introduce a symmetric matrix
$C^{IJ}_g$ through the relation,
%%%%%%%
\beq
\alpha^I_iC^{IJ}_g\alpha^J_j={1\over2}\alpha^I_i(B-U^TBU)^{IJ}
\alpha^J_j \quad {\rm mod} \ 2\ .
\eeq
%%%%%%%%%%%%%%%%
The right hand side is found to be
%%%%%%%%%
\beq
{1\over2}\alpha^I_i(B-U^TBU)^{IJ}\alpha^J_j=\pmatrix{0 & 1 \cr
                                                          -1 & 0
\cr}_{ij}\ ,
\eeq
%%%%%%%%%%%%%%%%
and hence $C^{IJ}_g$ cannot be chosen to be zero.
We may choose $C^{IJ}_g$ as
%%%%%%%%
\begin{eqnarray*}
\alpha^I_iC^{IJ}_g\alpha^J_j=\pmatrix{0 & 1 \cr
                                      1 & 0 \cr}_{ij}\ ,
\end{eqnarray*}
%%%%%%%%%%%%%
or
%%%%%%%%
\beq
C^{IJ}_g=\pmatrix{1 & 0 \cr
                  0 & -{1\over3} \cr}^{IJ}\ .
\eeq
%%%%%%%%%%%%%
This choice turns out to be consistent with $g^2=1$.

The action of $g$ on $(x^I_L,x^I_R)$ is anomalous and the action of
$g$ on the vertex operator is given by
%%%%%%%
\beq
g:\quad V(k_L,k_R;z)\ \rightarrow\
        e^{-i\frac{\pi}{2}(k_L-k_R)\cdot UC_gU^T(k_L-k_R)}
         V(U^Tk_L,U^Tk_R;z)\ .
\label{gV1}
\eeq
%%%%%%%%%%%%%%
Note that the nontrivial phase appears in the right hand side of
eq. (\ref{gV1}).
To see the importance of this phase, let us consider the following
operators of the left-movers:
%%%%%%%%
\beqs
H^I(z)&=&i\partial_zX^I_L(z)\ ,\nonumber\\
E^\alpha(z)&=&V(\alpha,0;z)\ ,
\eeqs
%%%%%%%%%%%%%%
where $\alpha$ is a root vector of $SU(3)$.
These operators form level one $SU(3)$ Ka{\v c}-Moody
algebra
\cite{Frenkel-KS}.
It turns out that the action of $g$ on the above generators
is given by
%%%%%%%%
\beqs
g:\quad
  H^I(z) &\rightarrow& U^{IJ}H^J(z)\ ,\nonumber\\
  E^{\pm\alpha_1}(z) &\rightarrow& E^{\pm\alpha_2}(z)\ ,\nonumber\\
  E^{\pm\alpha_2}(z) &\rightarrow& E^{\pm\alpha_1}(z)\ ,\nonumber\\
  E^{\pm(\alpha_1+\alpha_2)}(z) &\rightarrow&
   - E^{\pm(\alpha_1+\alpha_2)}(z)\ .
\eeqs
%%%%%%%%%%%%%%%%
Thus, the $Z_2$-invariant physical generators may be given by
%%%%%%
\beqs
J_3(z)  &=& 2H^1(z)\ ,\nonumber\\
J_\pm(z) &=& \sqrt{2}\left(
          E^{\pm\alpha_1}(z) + E^{\pm\alpha_2}(z)\right)\ .
\eeqs
%%%%%%%%%%%%%%%%
These generators are found to form level four $SU(2)$ Ka{\v c}-Moody
algebra
\cite{Sakamoto-T}.
It should be emphasized that the generators
$E^{\pm(\alpha_1+\alpha_2)}(z)$ are not invariant
under the action of $g$ and hence they are
removed from the physical generators although the root vector
$\alpha_1+\alpha_2$ is invariant under the action of $g$.

The zero mode eigenstates of the $g$-twisted sector will be
of the form,
%%%%%%
\beq
|\ k^1 + s^1_g , w^I\ >\ ,
\eeq
%%%%%%%%%%%%%%
where
%%%%%%%%
\beqs
k^1 &\in& 2\Lambda_W|_{_{/\!/}}
   = \{\ k^1 = \sqrt{2}m,\ m\in {\bf Z}\ \}\ ,\nonumber\\
w^I &\in& \frac{\Lambda_R}{(1-U)\Lambda_R}\ ,\nonumber\\
s^1_g &=& 1/\sqrt{2}\ .
\eeqs
%%%%%%%%%%%%%%
It may be convenient to rewrite $|k^1+s^1_g,w^I>$ as
%%%%%
\beq
|\ k^1_L+\frac{1}{2}s^1_g,k^1_R+\frac{1}{2}s^1_g; x^2\ >\
   \equiv\ |\ k^1+s^1_g, w^I\ >\ ,
\eeq
%%%%%%%%%%%%%%
where
%%%%%%%
\beq
(k^1_L,k^1_R)\ \in\ \Gamma^{2,2}|_{_{/\!/}}
   = \{\ (k^1_L,k^1_R) = (n/\sqrt{2},n'/\sqrt{2}),\ n,n' \in {\bf Z}\ \}\ .
\eeq
%%%%%%%%%%%%%%
Thus, the eigenvalues of $\widehat{p}^1$
(or $(\widehat{p}^1_L,\widehat{p}^1_R)$) are different from naively
expected values by $s^1_g$ (or $(s^1_g/2,s^1_g/2)$).
The number of the fixed points is given by
%%%%%%%
\beq
\frac{\det(1-\widetilde{U})}{V_{\Gamma_U}}
\ =\ \frac{2}{2}\ =\ 1\ ,
\eeq
%%%%%%%%%%%%%%%
where
%%%%%
\beq
\Gamma_U\ =\ (\Gamma^{2,2}|_{_{/\!/}})^{*}
        \ =\ \{\ (k^1_L,k^1_R) = (\sqrt{2}n,\sqrt{2}n'),\ n,n' \in
               {\bf Z}\ \}\ .
\eeq
%%%%%%%%%%%%%%%
It should be noted that the degeneracy of the ground state is different
from the number of the fixed points.
In fact, the ground state of the $g$-twisted sector
is fourfold degenerate, i.e.
%%%%%%%
\beq
|\pm\frac{1}{2}s^1_g,\pm\frac{1}{2}s^1_g;x^2=0\ >
\quad \mbox{and}\quad
|\pm\frac{1}{2}s^1_g,\mp\frac{1}{2}s^1_g;x^2=\frac{\pi}{2}
\sqrt{\frac{3}{2}}\ >\ .
\eeq
%%%%%%%%%%%%%%%

We finally discuss one loop modular invariance of the partition
function,
%%%%
\beq
Z(\tau)={1\over2}\sum_{\ell,m=0}^1Z(g^\ell,g^m;\tau)\ ,
\label{partitionZ2}
\eeq
%%%%%%%%%%%%%%
where
%%%%%%
\beqs
Z(1,1;\tau)&=&
   {1\over|\eta(\tau)|^4}\sum_{(k^I_L,k^I_R)\in\Gamma^{2,2}}e^{i\pi
   \tau (k^I_L)^2-i\pi{\bar\tau}(k^I_L)^2}\ ,\nonumber\\
Z(1,g;\tau)&=&
   \frac{|\vartheta_3(0|\tau)\vartheta_4(0|\tau)|}
        {|\eta(\tau)|^4}
   \sum_{(k^1_L,k^1_R)\in\Gamma_U}
     e^{i\pi(k^1_L-k^1_R)s^1_g}
     e^{i\pi\tau (k^1_L)^2-i\pi{\overline\tau}(k^1_R)^2}\ ,\nonumber\\
Z(g,1;\tau)&=&
   \frac{|\vartheta_3(0|\tau)\vartheta_2(0|\tau)|}
        {|\eta(\tau)|^4}
   \sum_{(k^1_L,k^1_R)\in{\Gamma_U}^{*}}
   e^{i\pi\tau(k^1_L+s^1_g/2)^2
      -i\pi{\overline\tau}(k^1_R+s^1_g/2)^2}\ ,\nonumber\\
Z(g,g;\tau)&=&
   \frac{|\vartheta_4(0|\tau)\vartheta_2(0|\tau)|}
        {|\eta(\tau)|^4}
   \sum_{(k^1_L,k^1_R)\in{\Gamma_U}^{*}}
   e^{i\pi(\tau+1)(k^1_L+s^1_g/2)^2
      -i\pi({\overline\tau}+1)(k^1_R+s^1_g/2)^2}\ .
\eeqs
%%%%%%%%%%%%%%%
The functions $\eta(\tau)$ and $\vartheta_a(\nu|\tau)$ $(a
=1,\cdots,4)$ are the Dedekind $\eta$-function and the Jacobi theta
function, which are defined in appendix B.
It is easily verified that $Z(g^\ell,g^m;\tau)$ satisfies the following
desired relations:
%%%%%%%
\beqs
Z(g^\ell,g^m;\tau+1)&=&Z(g^\ell,g^{m+\ell};\tau)\ ,\nonumber\\
Z(g^\ell,g^m;-1/\tau)&=&Z(g^{-m},g^\ell;\tau)\ ,
\eeqs
%%%%%%%%%%%%%%%%
and hence the partition function (\ref{partitionZ2}) is modular invariant.
It should be
emphasized that the existence of the shift vector
$(s^1_g/2,s^1_g/2)$ ensures
modular invariance of the partition function:
The left-right level matching condition (\ref{levelmatch'}), i.e.
%%%%%%
\beq
Z(g,1;\tau+2)=Z(g,1;\tau)\ ,
\eeq
%%%%%%%%%%%%%%%%
is satisfied because
%%%%%%%
\beq
2\Bigl(
      (k^1_L+\frac{1}{2}s^1_g)^2 - (k^1_R+\frac{1}{2}s^1_g)^2
 \Bigr) = 0\quad \mbox{mod}\ 2\ ,
\eeq
%%%%%%%%%%%%%%%
for all $(k^1_L,k^1_R) \in \Gamma^{2,2}|_{_{/\!/}}.$
If we put the shift vector $(s^1_g/2,s^1_g/2)$ to be zero by
hand, the level matching condition would be destroyed.

It is interesting to note that in terms of the theta functions the
partition function obtained above can be expressed as
%%%%%%
\beqs
Z(1,1;\tau)&=&{1\over|\eta(\tau)|^4}\sum_{(k^I_L,k^I_R)\in\Gamma^{2,2}}
e^{i\pi\tau (k^I_L)^2-i\pi{\overline\tau}(k^I_R)^2}\ ,\nonumber\\
Z(1,g;\tau)&=&{|\vartheta_3(0|\tau)\vartheta_4(0|\tau)|^2\over|\eta(
\tau)|^4}\ ,\nonumber\\
Z(g,1;\tau)&=&{|\vartheta_3(0|\tau)\vartheta_2(0|\tau)|^2\over|\eta(
\tau)|^4}\ ,\nonumber\\
Z(g,g;\tau)&=&{|\vartheta_4(0|\tau)\vartheta_2(0|\tau)|^2\over|\eta(
\tau)|^4}\ .
\eeqs
%%%%%%%%%%%%%%%%%%
This partition function is exactly identical to that of another
$Z'_2$ orbifold model whose $Z'_2$ transformation is defined by
%%%%%%%
\beq
Z'_2:\quad X^I\ \rightarrow\ -X^I\ ,
\label{Z2'}
\eeq
%%%%%%%%%%%%%%%%%
instead of the $Z_2$ transformation (\ref{Z2}). In this
$Z'_2$ orbifold model,
level one $SU(3)$ Ka{\v c}-Moody algebra can be shown to ``break'' to
level four $SU(2)$ Ka{\v c}-Moody algebra. Thus, although the two orbifold
models are defined by the different $Z_2$ transformations (\ref{Z2}) and
(\ref{Z2'}), they give the same spectrum and interaction
\cite{Sakamoto90}.

\vspace{1cm}

%%%%%%%%%%%%%%%%%%%%%%%%%%%%%%%%%%%%%%%%%%%%%%%%%%%%%%%%%%%%%%%%%%%
%%%%%%%%%%%%%%%%%%%%%%%%%%%%%%%%%%%%%%%%%%%%%%%%%%%%%%%%%%%%%%%%%%%

\sect{Discussion}

In this paper, we have discussed some of new aspects of string theory
on orbifolds in the presence of the antisymmetric constant background
field $B^{IJ}$ mainly in an operator formalism point of view.
The path integral formalism may be more suitable for discussing the
topological nature of the $B^{IJ}$-term.
In sect. 2, we have discussed the Euclidean path integral formalism for
string theory on {\it tori\/}
and have seen that the $B^{IJ}$-term can be
regarded as a topological term like a Chern-Simons term or a
Wess-Zumino term.
In the Euclidean path integral formalism for string theory on orbifolds,
the string coordinate $X^I(\sigma^1,\sigma^2)$ will in general
obey, instead of eq. (\ref{pathbc}), the following boundary conditions:
%%%%%%%% 9.1
\beqs
X^I(\sigma^1+\pi,\sigma^2)
   &=& U^{IJ}X^J(\sigma^1,\sigma^2) + \pi w^I\ ,\nonumber\\
X^I(\sigma^1,\sigma^2+\pi)
   &=& V^{IJ}X^J(\sigma^1,\sigma^2) + \pi w'^I\ ,
     \quad \mbox{for some}\ w^I,w'^I \in \Lambda\ ,
\eeqs
%%%%%%%%%%%%%%
with $[U,V]=0$.
For topologically trivial twists, i.e. $[U,B]=0=[V,B]$,
the Euclidean action is still given by eq. (\ref{eaction})
and the functional integral (\ref{epath}) directly gives the
expressions (\ref{pathint}) and (\ref{pathint'})
\cite{Inoue-NT}.
However, for topologically nontrivial twists,
a correct Euclidean action is not known yet and probably is different
from the naively expected one, i.e. eq. (\ref{eaction}).
The expressions (\ref{pathint}) and (\ref{pathint'})
strongly suggest that some additional
terms should appear in the action (\ref{eaction}).
To clarify the full topological nature of the $B^{IJ}$-term,
it is necessary to find a correct Euclidean action for topologically
nontrivial twists.

We have found Aharonov-Bohm like effects for twisted strings.
This result would seem strange because we might naively expect no
Aharonov-Bohm effect from eq. (\ref{CS}).
This is true for untwisted strings but
not for twisted strings.
Eq. (\ref{CS}) does not
imply no Aharonov-Bohm effect for twisted strings because
$S_E[X]$ in eq. (\ref{CS}) is the Euclidean action for strings
on tori.
The construction of a Euclidean action for twisted strings
will make clear geometrical meanings of the non-periodicity of wave
functions for twisted strings.

We have discussed the cocycle property of the vertex operators and
found that the zero modes of strings should obey the nontrivial
quantization conditions.
We have also found that a $Z_2$ phase appears in the transformation
(\ref{hV}).
The $Z_2$ phase (\ref{Z2phase}) depends on the winding
number $w^I=k^I_L-k^I_R$ and also the symmetric matrix $C^{IJ}_h,$
which is defined through the relation (\ref{Ch}).
Since $C^{IJ}_h$ is not defined uniquely in eq. (\ref{Ch}),
the twist operator $h$ would be specified by $h=(V;C_h)$.
It should  be noted that the group which consists of the
twist operators $h$'s is not necessarily equal to the group
which consists of the orthogonal matrices $V$'s.
In particular, $V^n=1$ does not necessarily mean $h^n=1$ but
$h^{2n}=1.$
Let $C'_h$ be another choice satisfying eq. (\ref{Ch}).
Then, we find
%%%%%%%%% 9.3
\beq
\frac{1}{2}w^I(C'_h-C_h)^{IJ}w^J = w^Iv^I\quad \mbox{mod}\ 2
\quad (w^I \in \Lambda)\ ,
\eeq
%%%%%%%%%%%%%%
for some constant vector $v^I.$
This means that the difference between $C'_h$ and $C_h$ can
be reinterpreted as the introduction of a $Z_2$-shift $v^I,$
i.e. $h'=(V;C'_h)=(V,v;C_h)$
and that the following two actions of $h'$ on $\widehat{x}^I$
and $\widehat{Q}^I$ give the same orbifold model:
%%%%%%
\beqs
h': \widehat{x}^I &\rightarrow& V^{IJ}\widehat{x}^J\ ,\nonumber\\
         \widehat{Q}^I &\rightarrow& V^{IJ}\widehat{Q}^J
         - [V,B]^{IJ}\widehat{x}^J +
         \pi V^{IJ}\Bigl( \frac{1}{2}(B-V^TBV) - C'_h \Bigr)^{JK}
         \widehat{w}^K\ ,
\eeqs
%%%%%%%%%%%%%%%
and
%%%%%%%
\beqs
h':\widehat{x}^I &\rightarrow& V^{IJ}\widehat{x}^J\ ,\nonumber\\
         \widehat{Q}^I &\rightarrow& V^{IJ}\widehat{Q}^J
         - [V,B]^{IJ}\widehat{x}^J +
         \pi V^{IJ}\Bigl(\Bigl( \frac{1}{2}(B-V^TBV) - C_h \Bigr)^{JK}
         \widehat{w}^K - v^J\Bigr).
\eeqs
%%%%%%%%%%%%%%%

%%%%%%%%%%%%%%%%%%%%%%%%%%%%%%%%%%%%%%%%%%%%%%%%%%%%%%%%%%%%%%%%%
%%%%%%%%%%%%%%%%%%%%%%%%%%%%%%%%%%%%%%%%%%%%%%%%%%%%%%%%%%%%%%%%%

\vspace{1cm}
\begin{center}
{\Large\bf Acknowledgements
  \par}
\end{center}

We should like to thank K. Ito, H. Nielsen, H. Ooguri, J. Petersen
and
F. Ruitz for useful discussions and also would like to acknowledge
the hospitality of the Niels Bohr Institute where this work was done.

\vspace{1cm}

%%%%%%%%%%%%%%%%%%%%%%%%%%%%%%%%%%%%%%%%%%%%%%%%%%%%%%%%%%%%%%%%%%%
%%%%%%%%%%%%%%%%%%%%%%%%%%%%%%%%%%%%%%%%%%%%%%%%%%%%%%%%%%%%%%%%%%%

\alphsec
\appe{A}

In this appendix, we derive eq. (\ref{cocycle}) with
eq. (\ref{cophase}).
We first give a precise definition of the vertex operator
$V(k_L,k_R;z)$ which describes the emission of an untwisted state with the
momentum $(k^I_L,k^I_R) \in \Gamma^{D,D},$
%%%%%%
\beqs
V(k_L,k_R;z) &\equiv&
 f_{k_L,k_R}(z)\ : \exp\{ik^I_LX^I_L(z)
                        + ik^I_RX^I_R(\overline{z})\} :\nonumber\\
&\equiv&
 f_{k_L,k_R}(z)\ \exp\{ik^I_LX^I_{L(-)}(z)
                        + ik^I_RX^I_{R(-)}(\overline{z})\}
               \    \exp\{ik^I_L\widehat{x}^I_L
                        + ik^I_R\widehat{x}^I_R\} \nonumber\\
& & \quad \times\: z^{k^a_L{\widehat p}^a_L}\: \overline{z}^
                   {k^a_R{\widehat p}^a_R}
               \  \exp\{ik^I_LX^I_{L(+)}(z)
                        + ik^I_RX^I_{R(+)}(\overline{z})\}\ ,
\eeqs
%%%%%%%%%%%%%%%%
where
%%%%%%%
\beqs
f_{k_L,k_R}(z) &=&
  \left(\frac{1}{z}\right)^{(k^\alpha_L)^2/2}
  \left(\frac{1}{\overline{z}}\right)^{(k^\alpha_R)^2/2}
  \prod^{N-1}_{m=1} (1+e^{-i\pi+i2\pi m/N})
    ^{-\frac{1}{2}k^I_L(1-U^m)^{IJ}k^J_L} \nonumber\\
& & \qquad \times \prod^{N-1}_{m=1}   (1+e^{i\pi-i2\pi m/N})
    ^{-\frac{1}{2}k^I_R(1-U^m)^{IJ}k^J_R}\ ,\nonumber\\
X^I_{L(-)}(z) &=& -i\sum_{n_J\in{\bf Z}-t_J/N>0}
   \frac{1}{n_J} M^{*IJ}\gamma^{J\dag}_{n_J} z^{n_J}\ ,\nonumber\\
X^I_{L(+)}(z) &=& i\sum_{n_J\in{\bf Z}-t_J/N>0}
   \frac{1}{n_J} M^{IJ}\gamma^{J}_{n_J} z^{-n_J}\ ,\nonumber\\
X^I_{R(-)}(\overline{z}) &=& -i\sum_{n_J\in{\bf Z}+t_J/N>0}
   \frac{1}{n_J} M^{*IJ}\overline{\gamma}^{J\dag}_{n_J}
     \overline{z}^{n_J}\ ,\nonumber\\
X^I_{R(+)}(\overline{z}) &=& i\sum_{n_J\in{\bf Z}+t_J/N>0}
   \frac{1}{n_J} M^{IJ}\overline{\gamma}^{J}_{n_J}
     \overline{z}^{-n_J}\ .
\eeqs
%%%%%%%%%%%%%%%5
In the following, we will use the commutation relations,
%%%%%%
\beqs
[\ \widehat{x}^I_L\ ,\ \widehat{p}^a_L\ ] &=& i \delta^{Ia}\ ,
   \nonumber\\
{[\ \widehat{x}^I_R\ ,\ \widehat{p}^a_R\ ]} &=& i \delta^{Ia}\ ,
   \nonumber\\
{[\ \gamma^I_{m_I}\ ,\ \gamma^{J\dag}_{n_J}\ ]} &=&
   m_I\delta^{IJ} \delta_{m_I,n_J}\ , \nonumber\\
{[\ \overline{\gamma}^I_{m_I}\ ,\ \overline{\gamma}^{J\dag}_{n_J}\ ]}
  &=& m_I\delta^{IJ} \delta_{m_I,n_J}\ .
\eeqs
%%%%%%%%%%%%%%%%%
Other commutation relations are equal to zero except for the
commutation relations between $\widehat{x}^I_L$ and $\widehat{x}^I_R,$
which will be derived in the text.
The vertex operator satisfies the following relations:
%%%%%%%%
\beqs
V(k_L,k_R;z)^{\dag} &=&
  \left(\frac{1}{\overline{z}}\right)^{(k^I_L)^2}
  \left(\frac{1}{z}\right)^{(k^I_R)^2}
  V(-k_L,-k_R;\frac{1}{\overline{z}})\ ,\nonumber\\
V(k_L,k_R;e^{2\pi i}z) &=&
  e^{-i\frac{\pi}{2}(k_L-k_R)^I(UC_gU^T)^{IJ}(k_L-k_R)^J}
  \ V(U^Tk_L,U^Tk_R;z)\ ,
\label{vertexrel}
\eeqs
%%%%%%%%%%%%%%%%%%
where we will use the identity (\ref{identity}) to derive the second
relation of eqs. (\ref{vertexrel}).
The vertex operator has the conformal weight
$((k^I_L)^2/2,(k^I_R)^2/2)$
and satisfies the following operator product expansions:
%%%%%%%%
\beqs
T(z') V(k_L,k_R;z) &=&
 \frac{(k^I_L)^2}{2(z'-z)^2} V(k_L,k_R;z) +
 \frac{1}{z'-z} \partial_z V(k_L,k_R;z) + \cdots, \nonumber\\
\overline{T}(\overline{z}') V(k_L,k_R;z) &=&
 \frac{(k^I_R)^2}{2(\overline{z}'-\overline{z})^2} V(k_L,k_R;z) +
 \frac{1}{\overline{z}'-\overline{z}}
  \partial_{\overline{z}} V(k_L,k_R;z) + \cdots.
\eeqs
%%%%%%%%%%%%%%%
It is not difficult to show that the operator product of
$V(k'_L,k'_R;z')$ and $V(k_L,k_R;z)$ is given by
%%%%%%
\beqs
\lefteqn{
 V(k'_L,k'_R;z')V(k_L,k_R;z)}  \nonumber\\
& & = f_{k'_L,k'_R}(z')\: f_{k_L,k_R}(z)\: \xi(k'_L,k'_R,z';k_L,k_R,z)
      \nonumber\\
& &\quad \times \exp\left\{ik'^I_LX^I_{L(-)}(z')
       + ik'^I_RX^I_{R(-)}(\overline{z}')
       + ik^I_LX^I_{L(-)}(z) + ik^I_RX^I_{R(-)}(\overline{z})\right\}
     \nonumber\\
& &\quad  \times V_0(k'_L,k'_R)\: V_0(k_L,k_R)\:
     {z'}^{k'^a_L{\widehat p}^a_L}\:
     {\overline{z}'}^{k'^a_R{\widehat p}^a_R}\:
     {z}^{k^a_L{\widehat p}^a_L}\: \overline{z}^{k^a_R{\widehat p}^a_R}
      \nonumber\\
& &\quad \times
     \exp\left\{ik'^I_LX^I_{L(+)}(z') + ik'^I_RX^I_{R(+)}(\overline{z}')
       + ik^I_LX^I_{L(+)}(z) + ik^I_RX^I_{R(+)}(\overline{z})\right\}\ ,
\eeqs
%%%%%%%%%%%%%
where
$V_0(k_L,k_R) = \exp\{ik^I_L\widehat{x}^I_L+ik^I_R\widehat{x}^I_R\}$
and
%%%%%%%%
\beqs
\lefteqn{\xi(k'_L,k'_R,z';k_L,k_R,z)}\nonumber\\
 & & =
    (z'-z)^{k'^I_Lk^I_L} (\overline{z}'-\overline{z})^{k'^I_Rk^I_R}
    \prod^{N-1}_{m=1}\left(z'^{1/N}+e^{-i\pi+i2\pi m/N}z^{1/N}\right)
     ^{-k'^I_L(1-U^m)^{IJ}k^J_L} \nonumber\\
 & &\quad \times \prod^{N-1}_{m=1}
      \left(\overline{z}'^{1/N}
            +e^{i\pi-i2\pi m/N}\overline{z}^{1/N}\right)
             ^{-k'^I_R(1-U^m)^{IJ}k^J_R}\ .
\eeqs
%%%%%%%%%%%%%%%%
The function $\xi(k'_L,k'_R,z';k_L,k_R,z)$ satisfies the following
relation:
%%%%%%%%
\beq
\xi(k'_L,k'_R,z';k_L,k_R;z) = \eta^{-1}\ \xi(k_L,k_R,z;k'_L,k'_R;z')\ ,
\eeq
%%%%%%%%%%%%%%%
where the phase $\eta$  is defined in eq. (\ref{cophase}).
As discussed in sect. 4, the operator product
$V(k'_L,k'_R;z')V(k_L,k_R;z)$ for $|z'|>|z|$ has to be analytically
continued to the region $|z|>|z'|$ and to be identical to
$V(k_L,k_R;z)V(k'_L,k'_R;z')$.
This requirement leads to the relation (\ref{cocycle})  with the
phase (\ref{cophase}).

\vspace{1cm}

%%%%%%%%%%%%%%%%%%%%%%%%%%%%%%%%%%%%%%%%%%%%%%%%%%%%%%%%%%%%%%%%%%%%
%%%%%%%%%%%%%%%%%%%%%%%%%%%%%%%%%%%%%%%%%%%%%%%%%%%%%%%%%%%%%%%%%%%%

\appe{B}

In this appendix, we present the definition of the theta functions
and various useful formulas which may be used
in the text.

We first introduce the theta function
%%%%%%%
\beq
\vartheta_{ab}(\nu|\tau)=\sum_{n=-\infty}^\infty \exp
    \left\{i\pi(n+a)^2\tau+i2
      \pi(n+a)(\nu+b)\right\}\ .
\eeq
%%%%%%%%%%%%%%
The four Jacobi theta functions are given by
%%%%%%%%%%%
\beqs
\vartheta_1(\nu|\tau) &=& \vartheta_{\frac{1}{2}\frac{1}{2}}(\nu|\tau),
   \nonumber\\
\vartheta_2(\nu|\tau) &=& \vartheta_{\frac{1}{2}0}(\nu|\tau),\nonumber\\
\vartheta_3(\nu|\tau) &=& \vartheta_{00}(\nu|\tau),\nonumber\\
\vartheta_4(\nu|\tau) &=& \vartheta_{0\frac{1}{2}}(\nu|\tau)\ .
\eeqs
%%%%%%%%%%%%%%%%%
They satisfy
%%%%%
\beqs
\vartheta_1(\nu+1|\tau) &=& -\vartheta_1(\nu|\tau)\ ,\nonumber\\
\vartheta_2(\nu+1|\tau) &=& -\vartheta_2(\nu|\tau)\ ,\nonumber\\
\vartheta_3(\nu+1|\tau) &=& \vartheta_3(\nu|\tau)\ ,\nonumber\\
\vartheta_4(\nu+1|\tau) &=& \vartheta_4(\nu|\tau)\ ,
\eeqs
%%%%%%%%%%%%%%
%%%%%%
\beqs
\vartheta_1(\nu+\tau|\tau)
  &=& -e^{-i\pi(\tau+2\nu)}\vartheta_1(\nu|\tau)\ ,\nonumber\\
\vartheta_2(\nu+\tau|\tau)
  &=& e^{-i\pi(\tau+2\nu)}\vartheta_2(\nu|\tau)\ ,\nonumber\\
\vartheta_3(\nu+\tau|\tau)
  &=& e^{-i\pi(\tau+2\nu)}\vartheta_3(\nu|\tau)\ ,\nonumber\\
\vartheta_4(\nu+\tau|\tau)
  &=& -e^{-i\pi(\tau+2\nu)}\vartheta_4(\nu|\tau)\ ,
\eeqs
%%%%%%%%%%%%%%%
%%%%%%
\beqs
\vartheta_1(\nu|\tau+1)
  &=& e^{i\frac{\pi}{4}}\vartheta_1(\nu|\tau)\ ,\nonumber\\
\vartheta_2(\nu|\tau+1)
  &=& e^{i\frac{\pi}{4}}\vartheta_2(\nu|\tau)\ ,\nonumber\\
\vartheta_3(\nu|\tau+1)
  &=& \vartheta_4(\nu|\tau)\ ,\nonumber\\
\vartheta_4(\nu|\tau+1)
  &=& \vartheta_3(\nu|\tau)\ ,
\eeqs
%%%%%%%%%%%%%%%
%%%%%%
\beqs
\vartheta_1(\nu/\tau|-1/\tau)
  &=& -i(-i\tau)^{1/2}e^{i\pi\nu^2/\tau}
      \vartheta_1(\nu|\tau)\ ,\nonumber\\
\vartheta_2(\nu/\tau|-1/\tau)
  &=& (-i\tau)^{1/2}e^{i\pi\nu^2/\tau}\vartheta_4
      (\nu|\tau)\ ,\nonumber\\
\vartheta_3(\nu/\tau|-1/\tau)
  &=& (-i\tau)^{1/2}e^{i\pi\nu^2/\tau}\vartheta_3
      (\nu|\tau)\ ,\nonumber\\
\vartheta_4(\nu/\tau|-1/\tau)
  &=& (-i\tau)^{1/2}e^{i\pi\nu^2/\tau}\vartheta_2
      (\nu|\tau)\ .
\eeqs
%%%%%%%%%%%%%%%%%
It is known that the Jacobi theta functions can be expanded as
%%%%%%%%%
\beqs
\vartheta_1(\nu|\tau)
  &=& -2q^{1/4}f(q)\sin\pi\nu\prod_{n=1}^\infty
      (1-2q^{2n}\cos2\pi\nu+q^{4n})\ ,\nonumber\\
\vartheta_2(\nu|\tau)
  &=& 2q^{1/4}f(q)\cos\pi\nu\prod_{n=1}^\infty(1
      +2q^{2n}\cos2\pi\nu+q^{4n})\ ,\nonumber\\
\vartheta_3(\nu|\tau)
  &=& f(q)\prod_{n=1}^\infty(1+2q^{2n-1}\cos2\pi\nu
      +q^{4n-2})\ ,\nonumber\\
\vartheta_4(\nu|\tau)
  &=& f(q)\prod_{n=1}^\infty(1-2q^{2n-1}\cos2\pi\nu
      +q^{4n-2})\ ,
\eeqs
%%%%%%%%%%%%%%%%
where
%%%%%%%
\beqs
q &=& e^{i\pi\tau}\ ,\nonumber\\
f(q) &=& \prod_{n=1}^\infty(1-q^{2n})\ .
\eeqs
%%%%%%%%%%%%%%%%
Another important function is the Dedekind $\eta$-function,
%%%%%%
\beq
\eta(\tau) = q^{1/12}\prod_{n=1}^\infty(1-q^{2n})\ ,
\eeq
%%%%%%%%%%%%%%%%
which satisfies
%%%%%
\beqs
\eta(\tau+1) &=& e^{i\frac{\pi}{12}}\eta(\tau)\ ,\nonumber\\
\eta(-1/\tau) &=& (-i\tau)^{1/2}\eta(\tau)\ .
\eeqs
%%%%%%%%%%%%%%%%%
We finally give the Poisson resummation formula, which will play a key
role in the modular transformation $\tau\rightarrow -1/\tau$. Let $
\Gamma^{d,\overline d}$ be a $(d+\overline{d})$-dimensional Lorentzian
lattice and ${
\Gamma^{d,\overline{d}}}{}^*$ be its dual lattice.
Then, the formula is given by
%%%%%%%
\beqs
\lefteqn{
 \sum_{(k_L,k_R)\in\Gamma^{d,\overline d}}
  e^{-i(\pi/\tau)(k_L+v_L)^2
      +i(\pi/\overline{\tau})(k_R+v_R)^2}
} \nonumber\\
& &\quad =
  \frac{(-i\tau)^{d/2}(i\overline{\tau})
   ^{\overline{d}/2}}{V_{\Gamma^{d,\overline{d}}}}
  \sum_{(q_L,q_R)\in\Gamma^{d,\overline{d}}{}^*}
  e^{i\pi\tau q_L^2-i\pi\overline{\tau} q^2_R+i2
  \pi(v_L\cdot q_L-v_R\cdot q_R)}\ ,
\eeqs
%%%%%%%%%%%%%%%
where $(v_L,v_R)$ is an arbitrary $(d+\overline{d})$-dimensional
constant vector
and $V_{\Gamma^{d,\overline{d}}}$ is
the volume of the unit cell of the lattice $\Gamma^{d,\overline{d}}$.

\vspace{1cm}

%%%%%%%%%%%%%%%%%%%%%%%%%%%%%%%%%%%%%%%%%%%%%%%%%%%%%%%%%%%%%%%%%%%%%
%%%%%%%%%%%%%%%%%%%%%%%%%%%%%%%%%%%%%%%%%%%%%%%%%%%%%%%%%%%%%%%%%%%%%

\appe{C}

In this appendix, we derive eqs. (\ref{pathint}) and (\ref{pathint'})
from eqs.
(\ref{partzero1}) and (\ref{partzero1'}).
To this end, let us first write eq. (\ref{partzero1})  as
%%%%%%%
\beq
Z(g,h;\tau) = \sum_{w^I\in \Lambda/(1-U)\Lambda} F(w^I)\ .
\label{c1}
\eeq
%%%%%%%%%%%%%%%%
Since $F(w^I)$ satisfies
%%%%%%%
\beq
F(w^I+(1-U)^{IJ}L^J) = F(w^I)\quad \mbox{for all}\ L^I \in \Lambda\ ,
\eeq
%%%%%%%%%%%%%%%%
we can rewrite eq. (\ref{c1}) as
%%%%%%
\beq
Z(g,h;\tau)_{zero} =
 \frac{1}{V_{\pi\Lambda_U}} \int_{\pi\Lambda} dx^I \sum_{w^I\in\Lambda}
 \delta\Bigl(x^\alpha-\pi\Bigl(\frac{1}{1-\widetilde{U}}\Bigr)
 ^{\alpha\beta}w^\beta\Bigr)\ F(w^I)\ ,
\label{c3}
\eeq
%%%%%%%%%%%%%%%
where $V_{\pi\Lambda_U}$  denotes
the volume of the unit cell of the lattice
$\pi\Lambda_U$ and
%%%%%%%
\beq
\Lambda_U = \{\ w^a\ |\ w^I=(w^a,w^\alpha=0) \in \Lambda\ \}\ .
\eeq
%%%%%%%%%%%%%%%%%
The integration of $x^I$ in eq. (\ref{c3}) runs over the volume
of the unit cell of the lattice $\pi\Lambda$, namely, over
the $D$-dimensional torus $T^D$.
Using the following identity:
%%%%%%
\beq
\int_{\pi\Lambda} dx^I \delta_{(1-\widetilde{V})^{ab}
      (k^b+s^b_g-B^{bI}w^I),0}\ f(x^\alpha)
= \int_{\pi\Lambda} dx^I e^{-i(k^a+s^a_g-B^{aI}w^I)
     (1-\widetilde{V})^{ab}x^b}\ f(x^\alpha)\ ,
\eeq
%%%%%%%%%%%%%%%%%
we can further rewrite $Z(g,h;\tau)_{zero}$ as
%%%%%%%
\beqs
\lefteqn{
Z(g,h;\tau)_{zero} }  \nonumber\\
& & = \frac{\pi^{d_2}}{V_{\pi\Lambda_U}\det(1-v_2)}
      \int_{\pi\Lambda} dx^I \int^{\infty}_{-\infty} dy^{i_2}
      \sum_{k^a\in 2\Lambda^{*}|_{_{/\!/}}}
      \sum_{w^I\in\Lambda}\sum_{w'^I\in\Lambda/\Lambda_U}
      \delta_{(1-V)^{IJ}w^J,(1-U)^{IJ}w'^J}\nonumber\\
& &\times
      \delta\Bigl(\pi\Bigl(\frac{1}{1-v^T_2}\Bigr)^{i_2j_2}y^{j_2}
      + \pi(k^{i_2}+s^{i_2}_g-B^{i_2I}w^I)\Bigr)\:
      \delta\Bigl(x^\alpha-\pi\Bigl(\frac{1}{1-\widetilde{U}}
       \Bigr)^{\alpha\beta}w^\beta\Bigr)\:
       \nonumber\\
& &\times
      \exp\Bigl\{
      i\pi\tau_1w^{i_1}(k^{i_1}+s^{i_1}_g-B^{i_1I}w^I)
      -\frac{\pi}{2}\tau_2\Bigl((k^{i_1}+s^{i_1}_g-B^{i_1I}w^I)^2
      + (w^{i_1})^2\Bigr) \nonumber\\
& &\quad
      + iy^{i_2}x^{i_2}-i\frac{\pi}{2}w^IC^{IJ}_hw^J
      -i\frac{\pi}{2}w'^IC^{IJ}_gw'^J
      +i\frac{\pi}{2}w'^I(U^TBV)^{IJ}w^J
      +i\frac{\pi}{2}w'^IB^{IJ}w^J \nonumber\\
& &\quad
      + i\pi w'^{i_1}(k^{i_1}+s^{i_1}_g-B^{i_1I}w^I)
      -i\pi w'^{i_2}\Bigl(\frac{1}{1-v^T_2}\Bigr)^{i_2j_2}y^{j_2}
      \Bigr\}\ ,
\eeqs
%%%%%%%%%%%%%%%%
where $\tau=\tau_1+i\tau_2$ and we have used the identity
%%%%%%%
\beqs
\lefteqn{e^{-i(k^a+s^a_g-B^{aI}w^I)(1-\widetilde{V})^{ab}x^b}}
     \nonumber\\
&=&
    \frac{\pi^{d_2}}{\det(1-v_2)}\int^{\infty}_{-\infty} dy^{i_2}
    e^{iy^{i_2}x^{i_2}}\:
    \delta\Bigl(\pi\Bigl(\frac{1}{1-v^T_2}\Bigr)^{i_2j_2}y^{i_2}
    + \pi(k^{i_2}+s^{i_2}_g-B^{i_2I}w^I)\Bigr).
\eeqs
%%%%%%%%%%%%%%%%%
Furthermore, using the formula,
%%%%%
\beqs
\lefteqn{
         \sum_{\Delta w^a=(\Delta w^{i_1},\Delta w^{i_2})\in\Gamma}
        \exp\Bigl\{-\lambda\sum^{d_1}_{i_1=1}
          (\Delta w^{i_1}+s^{i_1})^2
          +i\sum^{d_2}_{i_2=1}
         t^{i_2} \Delta w^{i_2}\Bigr\}  }  \nonumber\\
& & \qquad\qquad =
    \frac{(2\pi)^{d_2}}{V_\Gamma}
    \left(\frac{\pi}{\lambda}\right)^{d_1/2}
    \sum_{k^a=(k^{i_1},k^{i_2})\in 2\Gamma^{*}}
    \delta(t^{i_2}-\pi k^{i_2}) \nonumber\\
& &\qquad \qquad \qquad \times
    \exp\Bigl\{-\frac{\pi^2}{4\lambda}\sum^{d_1}_{i_1=1}(k^{i_1})^2
     +i\pi\sum^{d_1}_{i_1=1}k^{i_1}s^{i_1}\Bigr\}\ ,
\eeqs
%%%%%%%%%%%%%%%%%%
and integrating over the variables $y^{i_2},$ we have
%%%%%%%
\beqs
Z(g,h;\tau)_{zero}
&=&
 \frac{1}{\pi^{d_1}\det(1-v_2)}\left(\frac{1}{2\tau_2}\right)^{d_1/2}
 \int_{\pi\Lambda}dx^I \sum_{w^I\in\Lambda}
 \sum_{w'^I\in\Lambda/\Lambda_U}
 \sum_{\Delta w^I\in\Lambda_U}   \nonumber\\
& &\qquad \times\
 {\widetilde{\cal A}}(g,h;\tau;x^I,w^I,w'^I+\Delta w^I)\ ,
\eeqs
%%%%%%%%%%%%%%%%%%%
where we have used the fact that
$(\Lambda^{*}|_{_{/\!/}})^{*} = \Lambda_U.$
Replacing
%%%%%%%
\beqs
w'^I+\Delta w^I  &\rightarrow& w'^I\ ,\nonumber\\
\sum_{w'^I\in\Lambda/\Lambda_U} \sum_{\Delta w^I\in\Lambda_U}
&\rightarrow& \sum_{w'^I\in\Lambda}\ ,
\eeqs
%%%%%%%%%%%%%%%%%%%%
we finally obtain the expressions (\ref{pathint}) and (\ref{pathint'}).

%%%%%%%%%%%%%%%%%%%%%%%%%%%%%%%%%%%%%%%%%%%%%%%%%%%%%%%%%%%%%%%%%%%%%%
%%%%%%%%%%%%%%%%%%%%%%%%%%%%%%%%%%%%%%%%%%%%%%%%%%%%%%%%%%%%%%%%%%%%%%

\end{document}